\documentclass{aa}  
\usepackage{graphicx}
\usepackage{txfonts}
\usepackage{lscape}
\usepackage{graphicx}
\usepackage{epstopdf}
\usepackage{longtable}
\usepackage{supertabular}
\usepackage{natbib}
\bibpunct{(}{)}{;}{a}{}{,}
\def\r14{$R^{1/4}$}

\def\kmsMpc{km\,s$^{-1}$\,Mpc$^{-1}$}

\def\mueff{\ifmmode{\mu_{\rm e}}\else{$\mu_{\rm e}$}\fi}
\def\zf{\ifmmode{z_{\rm f}}\else{$z_{\rm f}$}\fi}
\def\Ks{\ifmmode{K_{\rm s}}\else{$K_{\rm s}$}\fi}

\renewcommand{\deg}{\ensuremath{{^\circ}}}

\newcommand{\VVmax}{\ensuremath{V/V_\mathrm{max}}}

\newcommand{\SExtractor}{\textsc{SExtractor}}
\newcommand{\Hyperz}{\textsc{HyperZ}}
\newcommand{\GimTwoD}{\textsc{Gim2D}}

\newcommand{\zspec}{\ensuremath{{z_\mathrm{spec}}}}
\newcommand{\zphot}{\ensuremath{{z_\mathrm{phot}}}}

\newcommand{\zform}{\ensuremath{{z_\mathrm{f}}}}
\newcommand{\Reff}{\ensuremath{{R_\mathrm{eff}}}}
\newcommand{\RdeV}{\ensuremath{R^{1/4}}}

\begin{document}
   \title{Bulges of disk galaxies at intermediate redshifts. I.}

   \subtitle{Samples with and without bulges in the Groth Strip Survey}

   \author{L. Dom\'\i nguez-Palmero
          \inst{1}
          \and
          M.~Balcells
          \inst{1}
          \and
          P.~Erwin
          \inst{2}
          \and
          M.~Prieto
          \inst{1}\inst{,3}
          \and
          D.~Crist\'obal-Hornillos
          \inst{1}\inst{,4}
          \and
          M.~C.~Eliche-Moral
          \inst{1}\inst{,5}
          \and
          R. Guzm\'an
          \inst{6}
}

   \offprints{L. Dom\'\i nguez-Palmero}

   \institute{Instituto de Astrof\'\i sica de Canarias, 
              E-38200 La Laguna, Tenerife, Spain \\
             \email{ldp@iac.es} 
              \email{balcells@iac.es} 
        \and
              Max-Planck-Institut fuer extraterrestrische Physik, 
              Giessenbachstrasse, D-85748 Garching, Germany
         \and
         	Departamento de Astrof\'\i sica, Universidad de La Laguna, 
		La Laguna, E-38200 Tenerife, Spain
         \and
              Instituto de Astrof\'\i sica de Andaluc\'\i a,
              18008 Granada, Spain 
         \and
              Departamento de Astrof\'\i sica y Ciencias de la Atm\'osfera, 
              Facultad de C.C.~F\'\i sicas, 
              Universidad Complutense de Madrid, E-28040 Madrid, Spain
          \and
          	Department of Astronomy, University of Florida, 
		211 Bryant Space Science Center, Gainesville, FL 32611-2055, USA
}

\date{\large \textit{Draft \today}}


 
  \abstract
   {Analysis of bulges to redshifts of up to $z \sim 1$ have provided ambiguous results as to whether bulges as a class are old structures akin to elliptical galaxies or younger products of the evolution of their host disks. }
   {We aim to define a sample of intermediate-$z$ disk galaxies harbouring central bulges, 
   and a complementary sample of disk galaxies without measurable bulges. We intend to provide colour profiles for both samples, as well as measurements of nuclear, disk, and global colours, which may be used to constrain the relative ages of bulges and disks. 
}
   {We select a diameter-limited sample of galaxies in images from the HST/WFPC2 (Wide-Field Planetary Camera 2 at the Hubble Space Telescope) Groth Strip survey, which is divided into two subsamples of higher and lower inclination to assess the role of dust in the measures quantities. 
Mergers are visually identified and excluded. We take special care to control the pollution by ellipticals. The bulge sample is defined with a criterion based on nuclear surface brightness excess over the inward extrapolation of the exponential law fitted to the outer regions of the galaxies. We extract colour profiles on the semi-minor axis least affected by dust in the disk, and measure nuclear colours at 0.85 kpc from the centre over those profiles. Disk colours are measured on major axis profiles; global colours are obtained from 2.6\arcsec\-diameter apertures. Colour transformations and K-corrections are calculated using SEDs covering bands $U B V I J K$, from the GOYA photometric survey. }
   {We obtain a parent sample containing 248 galaxies with known redshifts, spectroscopic or photometric, spanning $0.1 < z < 1.2$. The bulge subsample comprises 54 galaxies
($21.8\%$ of the total), while the subsample with no measureable bulges is $55.2\%$ of the total (137 galaxies).  The remainder ($23\%$) is composed of mergers.  We list nuclear, disk, and global colours (observed and rest-frame) and magnitudes (apparent and absolute), as well as galaxy colour gradients for the samples with and without bulges, and make them available in electronic form{\thanks{Tables~\ref{tab:bul_datos} to~\ref{tab:nobul_gradient} are available in electronic form at the CDS via anonymous ftp to \textsf{cdsarc.u-strasbg.fr (130.79.128.5)} {\rm or via} \textsf{http://cdsweb.u-strasbg.fr/cgi-bin/qcat?J/A+A/}}}. We also provide images, colour maps, plots of spectral energy distributions, major-axis surface brightness profiles, and minor-axis colour profiles for both samples.}
   {}


   \authorrunning{Dom\'\i nguez-Palmero et al.}
   \titlerunning{Bulges of disk galaxies at intermediate redshifts. I.}

   \maketitle
%

\section{Introduction}
\label{sec:introduction}

Bulges of spiral galaxies are central ingredients in the study of galaxy formation.  Traditionally seen as elliptical galaxies that happen to live inside disks \citep{Renzini99}, bulges are still often used by modellers as yardsticks of the importance of violent processes vs gentle accretion in the mass assembly of galaxies \citep[e.g.,][]{Kauffmann96bul, Cole00, Saiz01}. In the now termed 'classical' hypothesis,  bulges form before disks, at high redshifts, via major mergers \citep{Baugh98} or primordial collapse \citep{Eggen62}. Subsequently, a new disk forms out of the remaining gas. However, other hypotheses have been presented for bulge formation \citep[see][for two recent reviews]{Wyse97,Kormendy04}.  In a second scenario, bulges form after disks through bar instabilities \citep{Pfenniger90}, clumpy fragmentation of the existing disk \citep{Noguchi00,Immeli04}, or other secular processes; see \citet{Athanassoula05} for a discussion of the relation of these processes to disky and peanut-shaped bulges.  A related model posits that bulges may be the product of initial star formation in a galaxy, which proceeds until the bulge becomes massive enough to allow subsequent incoming gas to form a stable disk \citep{vandenBosch98}.
Minor mergers probably have a role in the growth of bulges and disks \citep{Aguerri01,Abadi03,Bournaud05,elichebulges06}, leading perhaps to a coeval growth of both components.  

Knowing the relative ages of bulges and disks is key to being able to falsify one or more of the hypothesis for bulge formation.  A prevalence of blue bulges would argue against an old formation age for bulges.  In the disk instability model, blue, starbursting bulges would need to be found in sufficient quantities to account for the mass growth of today's massive, red bulges. The third scenario would be supported by the finding of similar colour for bulges and disks.

Usually, two approaches are followed for obtaining bulge ages and exploring  mechanisms of bulge formation and evolution. The first is the study of kinematic, luminosities, colours, and chemical abundances of local bulges to recover the tracks of their evolution. The second is the analysis of global properties of galaxies at high redshifts with the aim of studying their evolution nearer the epoch at which bulges may have formed.

In the local Universe, \citet{Peletier96} found that the age differences between bulges and disks was, in most cases, consistent with zero.  Similarly, \citet{Bouwens99} concluded that dating uncertainties prevent us from distinguishing between the various bulge formation models. Indeed, age dating of stellar populations has large uncertainties. In the present series of papers, we address the \citet{Peletier96} study at redshift up to $z \sim 1$. 

To study bulges at high redshift, we need large samples from deep images at high spatial resolution. Deep surveys with HST provide enough resolution to resolve big bulges at cosmological distances; e.g. at $z=1$, a bulge with a size of 2 kpc subtends 0.26\arcsec\, about twice the point spread function (PSF) of the HST/WFPC2. But defining and characterising bulge components faces several practical difficulties derived from the properties of the imaging survey and the adopted selection criteria. First, pixelation and convolution with the PSF make the detection of small bulges difficult. Second, cosmological surface brightness dimming prevents detection of extended disks. Hence, samples selected by magnitude will favor compact, spheroidal galaxies, while diameter-limited selection will include only those bulges embedded in moderate- or high-surface brightness disks. Bulge-disk decompositions of the surface brightness profiles, already difficult in the local Universe, are made harder at high redshifts because bulges are not mapped with enough resolution and also sometimes because of the irregular morphologies of many disks. Moreover, the bulge colours based on bulge-disk decomposition have the additional problem of assigning a single representative colour for the bulges and for the disks. As the colour of the disk is usually not uniform, one can substract from bulge region disk light that is too blue or to red, which biasses the bulge colour.

Many recent studies of high-redshift galaxies are relevant to the formation of bulges \citep[e.g.,][to name just a few]{Marleau98, Menanteau01, Ellis01, Hammer01, Simard02, Conselice04, Koo05, Koo05HDF, Elmegreen07,MacArthur07}.  The range of agreement and discrepancies between the conclusions of these studies is strongly affected by differences in sample selection and in the methodology to infer age information from the data.  
In the present paper, we approach the definition of bulge samples at high redshift in a complementary way to previous studies.  
The aim is to work with samples of disk galaxies, i.e., we focus on bulges residing inside disks.  We then isolate samples with and without measurable bulges, obtain colour profiles, and measure nuclear, disk and global colours. Specifically, we work with a diameter-limited sample in the Groth Strip survey, divided in  two subsamples by inclination. From those samples we exclude mergers (visually identified by their morphology), and take care to control sample pollution by bonafide ellipticals, which should be minimised in the high-inclination sample. We define a prominence index $\eta$ as the surface brightness excess over the inward extrapolation of the outer disk profile, and use it to identify the sample of galaxies with measurable bulges. As a result, we get the sample of intermediate-$z$ bulge galaxies, but also we obtain a comparison sample of disk galaxies with no measurable bulges. We provide nuclear, disk, and global colours for galaxies in both samples in tabular form.  

Details on the approach followed here are given in Sect.~\ref{sec:approach}.  Section~\ref{sec:data} describes the HST and ground-based data. In Sect.~\ref{sec:sample}, we describe the characteristics of the high- and low-inclination samples, compare different methods to select bulge sample and describe our bulge sample selection method, discuss completeness issues as well as comparison samples at intermediate redshfits in the same survey. Bulge and disk colour measurements and K-corrections are detailed in Sect.~\ref{sec:methodology}. The characteristics and colours of different galaxy samples with and without bulges are tabulated in Appendix~\ref{sec:datatables}; in Appendix~\ref{sec:samplesbulges}, we show the HST/WFPC2 $F814W$ images, also colour maps, semi-major axis surface brightness profiles and semi-minor axis colour profiles of the bulge sample.

In a second paper \citep[hereafter Paper~II]{Dominguez08II}, we will analyse the nuclear, disk, and global colours of the galaxies presented here, establish relations between different parameters, and discuss implications of the results for galaxy formation and evolution models. A cosmology with  $\Omega_M = 0.3$, $\Omega_\Lambda = 0.7$, $H_0 = 70$ \kmsMpc\ is assumed throughout.  Magnitudes are expressed in the Vega system.  

\section{Approach}
\label{sec:approach}

The primary focus of the paper is on bulges residing in bonafide disk galaxies.  While there is sound evidence to assume strong similarities between (massive) bulges and ellipticals in the local Universe, we want to protect our high-redshift analysis from the biasses derived from indiscriminately polluting our bulge samples with ellipticals.  

Second, we want to control the effects of inclination on the colour measurements. For bonafide bulges, i.e., those that protrude above the disk, and given sufficient spatial resolution, an inclined aspect is the safest aspect angle to measure colours in bulges. While inclined aspects are bad for disks because they enhance dust reddening, inclined views are superior for bulges because half the bulge is seen above the disk, where most of the dust resides. Measuring colours in this half and discarding the other one, we get reliable bulge colours with minimised dust reddening.  This approach was used by \citet{Balcells94} to define and provide colours for a reference sample of bulges in the local Universe. When applying this method to determine bulge colours at intermediate redshift, we face two limitations.  First, when colour measurements are performed at galactocentric distances comparable to the size of the PSF, light from both sides of the centre gets convolved, and we risk losing the dust-free advantage provided by the inclined aspect angle.  Second, when the objects we measure do not correspond to bona-fide bulges, but rather to those inner brightened disks that \citet{Kormendy93} named pseudobulges, then the inclined view is actually more prone to dust reddening than the face-on views.  Given these difficulties, we design our experiment in a way that allows us to detect any effects of dust reddening, if present.  Hence, we divide the initial parent sample in two subsamples with low ($< 50\deg$) and high inclinations ($> 50\deg$) and analyse them independently. Keeping the high- and low-inclination samples separate has the added advantage that sample polution by ellipticals is almost non-existent in the high-inclination sample (see Sect.~ \ref{sec:ellipticalcontamination}).

As our third starting point we work with direct colour measurements, rather than colour measurements derived from a bulge-disk decomposition of the galaxy images.  
Direct colour measurements suffer from disk contamination.  But, in practice, colour gradients in disk galaxies are shallow, both in the local Universe \citep{Peletier96} and at cosmologically-significant redshifts.  Substraction of disk light allows one to target fainter bulges but the dependence of the results on the bulge-disk modelling is difficult to predict, especially for high redshift galaxies which often depart from axisymmetry. Accounting for the colour gradients when modelling the profiles in two bands is challenging, hence it is common practice to assume uniform colours for the disk and for the bulge. However, in the common case of a global negative colour gradient, this will artificially redden the bulge colours. In our approach, we take advantage of the observation that galaxy colour profiles and colour maps show very small negative gradients in all semi-minor and semi-major axes. We therefore infer that the colours of the inner regions of the disk are very similar to those of the bulge and we do not correct our colour measurements from disk contamination.
We provide checks on the amount of disk effects on the bulge colours by modelling the disk contribution under the assumption of uniform disk colours.

We finally assume that there is no unique way to correctly define a sample of bulges at high redshift.  If our goal is to learn the ages of present-day bulges, we would need to include not only galaxies that look like bulges, but also all those galaxies that will grow bonafide bulges at lower redshifts than $z_\mathrm{obs}$.  Otherwise we will be biassing the sample of progenitors of present-day bulges in favor of the oldest progenitors \citep{vanDokkum01}.  While a full treatment of such bias is beyond the scope of this work, we simply look for central light concentration as an indication that a bonafide bulge exists or may exist in the future.  Using the observation that many disk galaxies show exponential surface brightness profiles \citep{Freeman70}, we define a bulge prominence index $\eta$ that measures the excess central surface brightness over the inward extrapolation of the outer exponential profile and use it to select a sample of intermediate-redshift bulges.  Such an index is a sort of concentration index, but is more sensitive to small bulges in extended disks than standard concentration indices \citep{Abraham96,Conselice03} with which we compare the resulting selections.  

The distance bias affecting the $\eta$ index is similar to that of bulge-disk decompositions -- small bulges are progressively missed as we move to higher $z$.  We choose not to attempt any correction of that bias.  We provide measurements for any bulge that is resolved by the HST/WFPC2 images, which is over two resolution elements.  Distance and resolution effects need to be accounted for as part of the analysis of the results.  

\section{Data}
\label{sec:data}

\subsection{HST imaging}
\label{sec:HSTimaging}
 
Our intermediate-$z$ galaxy samples were selected from the HST imaging of the  Groth-Wesphal Strip \citep[GWS; see, e.g.,][]{Groth94,Roche96}.  The GWS was included in the Medium-Deep Survey \citep[MDS; see, e.g.][]{Ratnatunga99}, and has been one of the main targets of the Deep Extragalactic Evolutionary Probe \citep[DEEP; see, e.g.,][]{Simard02}.   The original HST survey consists of 28 overlapping WFPC2 subfields covering a total area of 113 arcmin$^{2}$ in a $45'$-long strip oriented NE to SW, with a position angle of $40\deg 3' 48\arcsec$  centred in RA(J2000) = $14^{\rm h} 16^{\rm m} 38.8^{\rm s}$, DEC(J2000) = $52\deg 16' 52.32\arcsec$, at galactic latitude $b \thicksim 60\deg$. For 27 of those fields, exposure times are 2800 s in the broad $V$ filter $(F606W)$ and 4400 s in the broad $I$ filter $(F814W)$. Field number 28, the fourth from the NW (Westphal Deep Field), has total exposures of 24400 s in $F606W$ and 25200 s in $F814W$. 

The GWS HST data were obtained from the HST archive and were re-calibrated on-the-fly with the best reference files available at the CADC (Canadian Astronomy Data Center)\footnote{\textsf{http://cadcwww.hia.nrc.ca/}}. The calibration includes: flagging of static bad pixels, analog-to-digital (A/D) correction, subtraction of bias level, subtraction of bias image, subtraction of dark for exposures longer than 10 s, flatfield correction, shutter shading correction to exposures of less than 10 s, and computation of photometry keywords.

After that, we removed cosmic rays using the \textsc{crrej} task from the {\textsc{stsdas.hst-calib.wfpc}} calibration package, and also corrected warm pixels in the images using the IRAF task \textsc{cosmicrays}, which replaces a detected warm pixel with the average of the four closest neighbors.

\subsection{Ground-based optical-NIR imaging}
\label{sec:GBimaging}

We obtained ground-based imaging of the GWS from the GOYA (Galaxy Origins and Young Assembly) photometric survey. 
The GOYA photometric survey is a near-infrarred (NIR) and optical survey in six broadband filters 
($U$, $B$, $F606W$, $F814W$, $J$, \Ks) covering, amongst other fields, the Groth strip, with target depths of $U$=$B$=$F606W$=$F814W$=26, and $J$=\Ks=22 (AB mag).  These data are taken in preparation for the GOYA project \citep{Guzman03,Balcells03}. GOYA is designed to exploit the NIR multi-object capabilities of the EMIR cryogenic spectrograph now under construction for the 10m GTC \citep[see][]{Balcells98,Balcells00spie,Garzon05,Garzon06}, to carry out an extended census of the optical rest-frame spectra of $1<z<3$ galaxies.  

The $J$- and \Ks-band data for the Groth strip  consist of dithered pointings with 96 min exposures in \Ks\ and 32 min in $J$, with the INGRID Hawaii-1 camera on the William Herschel telescope (WHT) at the Observatorio Roque de los Muchachos (ORM). The $50\%$ detection efficiencies range between $\Ks\ = 21.24$ mag and $\Ks\ = 20.21$ mag, depending on the seeing of the individual pointings. Details of the data reduction may be found in \citet{Cristobal03}.  $U$ and $B$ photometry are described in \citet{Eliche06}.  They were obtained with the WFC camera on the prime focus camera of the Isaac Newton Telescope (INT) at the ORM.  Detection depths at the $50\%$ efficiency were 24.83 mag and 25.45 mag in $U$ and $B$, respectively.  

With the combined HST and ground-based data, we derived broadband spectra energy distributions (SEDs) via the \SExtractor\ package, using 2.6\arcsec circular apertures on images smoothed to a common 1.3\arcsec\ FWHM PSF (see Appendix~\ref{sec:samplesbulges}).

We selected best-fit SED templates for each source from those in the {\textsc {Hyperz}} package \citep{hyperz}. We used such templates for photometric redshifts (Sect.~\ref{sec:redshifts}) and K-corrections (Sect.~\ref{sec:completeness} and Sect.~\ref{sec:colourtransformation}).  

\subsection{Redshifts}
\label{sec:redshifts}

More than $50\%$ of the sources selected for this work (131 galaxies) have spectroscopic redshifts publically available as part of the DEEP1 data release of the DEEP Groth Strip Survey (GSS)\footnote{\textsf{http://deep.ucolick.org/}}; see \citet{Weiner05} for details on the determination of the redshifts. For the remainder of the objects, 117 galaxies, we derived photometric redshifts using \Hyperz\ v1.2. Photometry in the $U$, $B$, $F606W$, $F814W$, $J$, and \Ks\ bands were fitted with template spectral energy distributions, using DEEP1 spectroscopic redshifts as a training set.  The solutions show fractional errors $\langle|\Delta(z)|/(1+z)\rangle = 0.071$.  

\begin{figure}[!htb]
\begin{center}
\includegraphics[angle=0,width=0.45\textwidth]{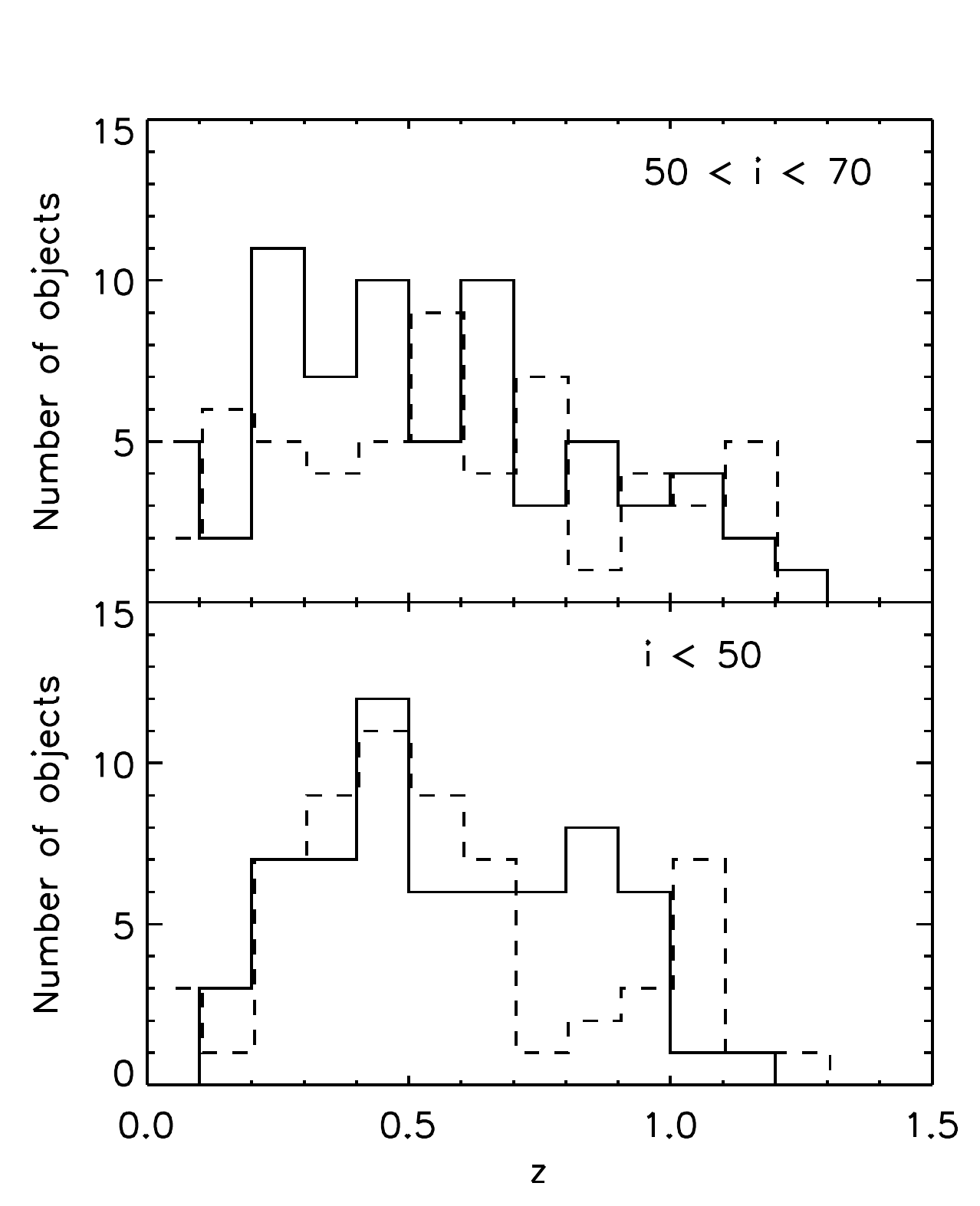}%
\end{center}
\caption{Distribution of redshifts for the high-inclination and low-inclination samples. {\it Solid line}: spectroscopic redshifts. {\it Dashed line}: 
photometric redshifts.}  \label{fig:distribredshift}
\end{figure}

\begin{figure}[!htb]
\begin{center}
\includegraphics[angle=0,width=0.45\textwidth]{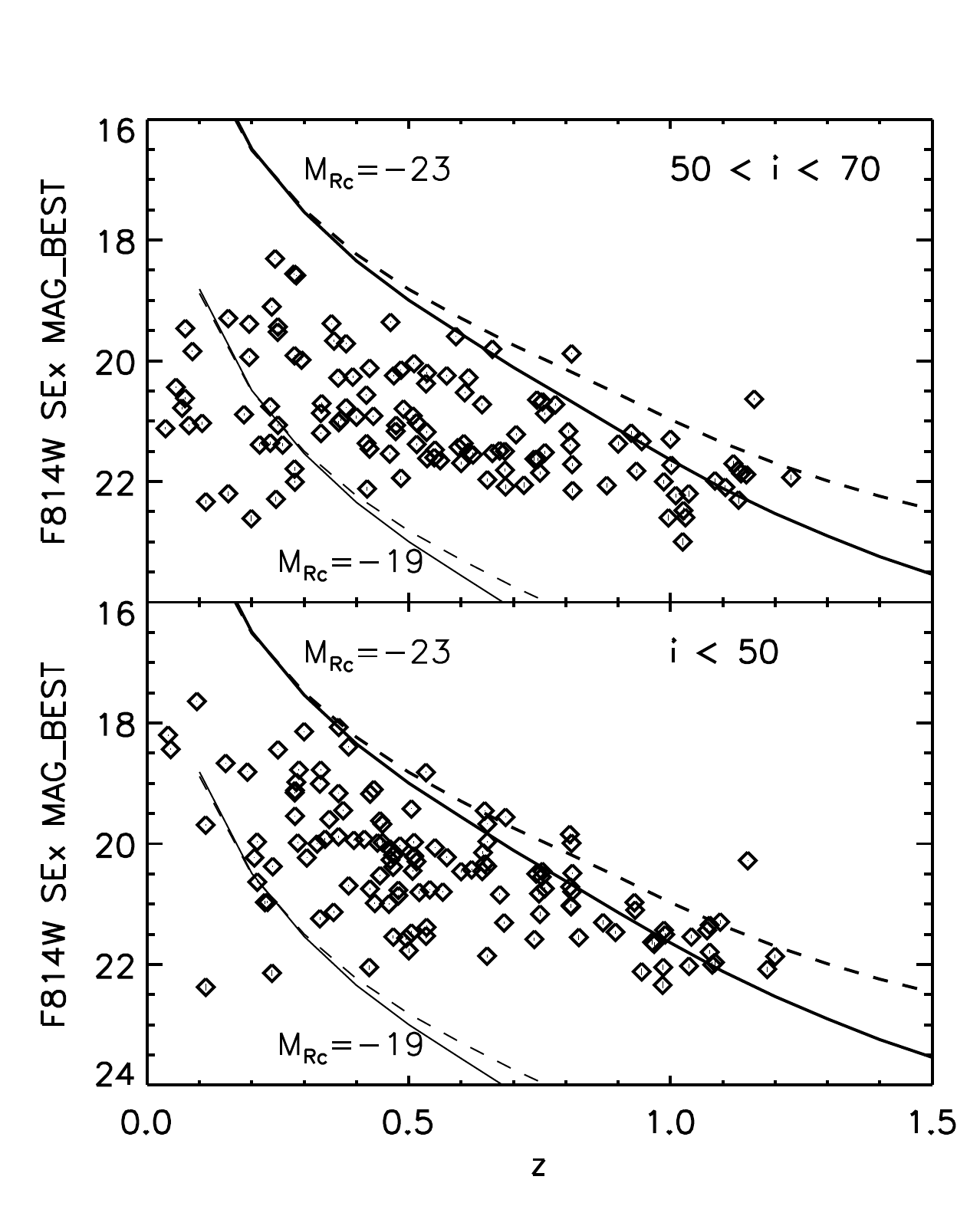}%
\end{center}
\caption{Distribution of $F814W$ apparent magnitude (\SExtractor\ MAG\_BEST magnitudes) vs redshift for the high-inclination and low-inclination samples. {\it Thick solid and dashed lines}: apparent magnitude of Sa and Sd galaxy models (from GISSEL03 library, \citet{Bruzual03}; \citet {Chabrier03} IMF, solar metallicity and exponential SFR with $\tau=3 and 30$ Gyr), respectively, for an $R$-band absolute magnitude equal to -23. {\it Thin lines}: the same but for $M_{R} = -19$.}  \label{fig:distribmagapI}
\end{figure}

\begin{figure}[!htb]
\begin{center}
\includegraphics[angle=0,width=0.45\textwidth]{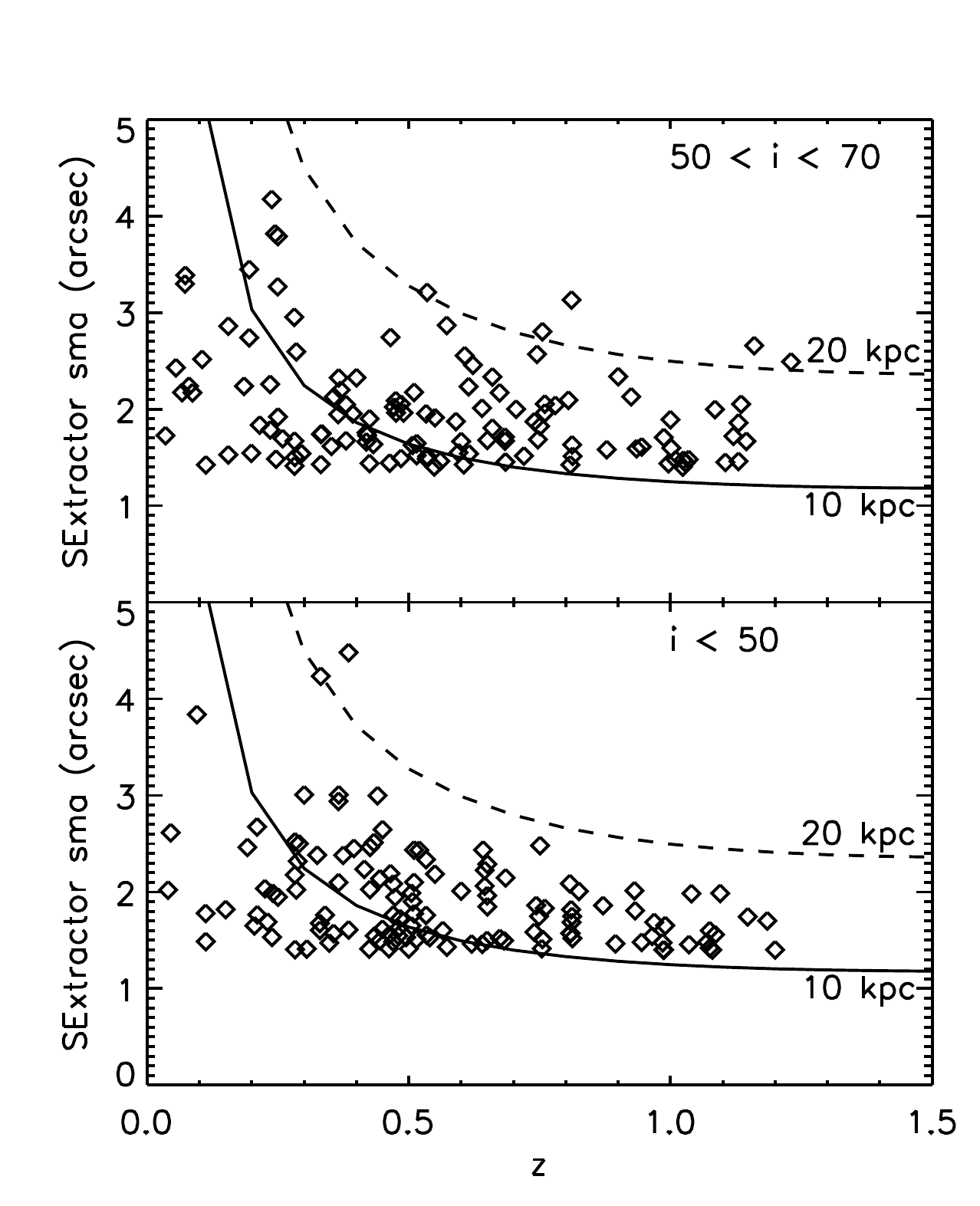}%
\end{center}
\caption{Distribution of $F814W$ semi-major axis in arcsec (from \SExtractor\ 'A-WORLD' size parameter) vs redshift for the high-inclination and low-inclination samples. {\it Solid line}: apparent size in arcsec for galaxies with a semi-major axis equal to 10 kpc. {\it Dashed lines}: the same for a semi-major axis equal to 20 kpc.}  \label{fig:magapI}
\end{figure}

\section{Sample Selection}
\label{sec:sample}

\subsection{Disk galaxy sample}
\label{sec:disksample}

Following the ideas outlined in Sect.~\ref{sec:approach}, we started by defining high-inclination and low-inclination samples on the GSS $F814W$ HST images. We divide the sample by inclination because we intend to study the possible effects of dust with inclination.
We produced object catalogs with \SExtractor\ version 2.3.2\footnote{\textsf{http://terapix.iap.fr/soft/sextractor}} \citep{Bertin96}, using a detection threshold of $1.5\sigma$ and a detection minimum area of 16 pixels. We selected all galaxies with apparent major-axis radii $R > 1.4$\arcsec. The value of R corresponds to the semi-major axis of the ellipse that describes the isophotal limit of a detected object, it is calculated as three times the maximum spatial $rms$ of the object profile along the semi-major axis, which is given by \SExtractor\ \texttt{A\_IMAGE} parameter.
Working on galaxies with apparent diameter above a given limit 
provides sufficient resolution elements to map the bulge and disk components,
and also helps to exclude compact, low-luminosity ellipticals from the sample. Our simulations show that $R > 1.4$\arcsec\ cut is a good matches for the apparent sizes that large spirals in the local Universe would have if placed at the typical redshifts of our study.  Additionally, a diameter-limited sample is easier to turn into a volume-limited sample, using \VVmax\ techniques (Sect.~\ref{sec:completeness}).  

The high-inclination sample was selected by applying a lower cut to the \SExtractor\ ellipticity of 0.37, which corresponds to inclinations above $50\deg$ for disks.  To avoid linear galaxies and edge-on disks, which pose specific problems for the data analysis, we excluded all galaxies with ellipticities above 0.66 (disk inclinations above $70\deg$). After excluding those galaxies too close to the image edges and those with biassed photometry, the resulting $50\deg<i<70\deg$ sample contains 170 galaxies. Of those, 123 have redshift information, either spectroscopic or photometric (Sect.~\ref{sec:redshifts}), and configure our parent inclined sample. 
The remaining 47 galaxies had no spectroscopic redshifts, and photometric redshifts were deemed unreliable due to lack of either $UB$ or NIR photometry.  Such cases were not included in the sample. In particular, there were 19 objects with no detection in $U$ or $B$, or in both, and 24 that lack photometry in $J$ or $K$, or in both. The remainder 4 galaxies belong to a field that was not observed in $J$.

As discussed in Sect.~\ref{sec:approach}, the high-inclination selection of this sample should filter out most ellipticals, leaving a bonafide disk galaxy sample. The low-inclination sample was derived by selecting all galaxies with inclinations $i < 50\deg$ from the $R > 1.4$\arcsec diameter-limited catalog.  This yielded 142 objects. Of those, the 125 objects with known spectroscopic or photometric redshifts comprise the parent low-inclination sample. 
The remaining 17 galaxies were not included, as was done with the high-inclination sample, due to incomplete photometric coverage. There were 7 objects that lack $UB$ photometry and 8 without data in NIR bands; 2 objects belong to the field not observed in $J$.
  
The redshift distributions of the high- and low-inclination samples are shown in Fig.~\ref{fig:distribredshift}.  Redshifts, spectroscopic and photometric, cover the range $0.1 < z < 1.3$, the median values are equal to $0.53$, for the high-inclination sample, and to $0.51$, for the low-inclination one; the two distributions are statistically similar to one another, with a $45\%$ probability of belonging to the same parent distribution, resulting from a K-S test.

Apparent magnitudes are in the range $18 < F814W < 23$ (Fig.~\ref{fig:distribmagapI}).  On the bright end, the upper envelope corresponds to Sa and Sd galaxy models with $M_{R} = -23$. The absolute magnitude corresponding to the faint end varies of course with redshift.  Both high- and low-inclination samples extend down to $M_{R} = -19$ for $z < 0.5$, i.e., the magnitude domain of dwarf galaxies. Additionally, Fig.~\ref{fig:distribmagapI} shows that, at low redshifts ($z< 0.3$), the high-inclination sample reaches down to fainter magnitudes ($M_{R} \sim -16$) than the low-inclination sample. This difference is due to the higher surface brightness of inclined galaxies, which yields higher apparent sizes.  These differences arise only for low-luminosity galaxies at low redshift, the dwarf domain we are not interested in. For brighter magnitudes, we see a marginal deficit of bright galaxies in the high-inclination sample when compared to the low-inclination one, which may be due to higher extinction in highly-inclined objects. It is also possible that the brightest low-inclined objects at each redshift are massive spheroidal galaxies.

The distribution of the \SExtractor\ apparent semi-major axes versus redshift is shown in Fig.~\ref{fig:magapI}. Solid and dashed lines represent the angular sizes corresponding to objects with physical sizes of 10 kpc and 20 kpc, respectively. Therefore, typical disk sizes are 10$-$15 kpc, except at $z<0.3$ where many smaller objects enter into the samples.  
As expected, bigger sizes are found in the high-inclination sample due to the effect of inclination in the optical path, but the difference is small and we decided not to correct this effect.

\subsection{Sample completeness}
\label{sec:completeness}

We use the \VVmax\ formalism to assess the statistical completeness of our parent samples.  To that end, we artificially shift each galaxy image over redshift, and determine the range of redshifts at which each galaxy would satisfy the size criteria we imposed on the original sample: $R > 1.4$\arcsec\, i.e., the maximum distance of the galaxy to be included in the sample. Cosmological dimming, K-corrections, background brigthness, pixelation, and image PSF are taken into account to mimic the $F814W$ HST/WFPC2 images of the Groth Strip, with K-corrections computed using best-fit SED templates \citep{Bruzual03} derived for each galaxy from \Hyperz\ best-fit solution.

To check that the sample is statistically complete, we have applied a $V/V_{\rm max}$ test \citep[]{Schmidt68,Thuan79}, where $V/V_{\rm max}$ is the ratio between the accesible volume of a galaxy and the maximum accesible volume for the same galaxy. For a randomly distributed sample of objects the mean value of $V/V_{\rm max}$ should be $0.5 \pm 1 / \sqrt{12 \times N}$, where $N$ is the number of objects in the test. The average value we have obtained is $V/V_{\rm max} = 0.41 \pm 0.03$ and $V/V_{\rm max} = 0.42 \pm 0.03$, respectively, for the high- and low-inclination samples; thus, the sample has a mild bias toward nearby objects.  

\subsection{Galaxies with measurable bulges}
\label{sec:bulsample}

In identifying and characterising bulge components in our parent galaxy sample, we face several practical difficulties. The first is spatial resolution:  with HST/WFPC2, one 0.1\arcsec\ pixel projects onto 0.8 kpc at $z \sim 1$.  As a result, many local bulges would become hidden within the central pixel if they were viewed at redshifts approaching $z=1$.  Second, we preferred to subtract the contribution from the disk, but bulge-disk decomposition, already an uncertain industry at low redshifts,  is made difficult by the irregular morphologies of many high-$z$ disks. 

Given these difficulties, we tried five different methods and compared them to one another before adopting our bulge selection criterion. 
First, we carried out a bulge-disk decomposition with the two-dimensional fitting code of \citet{Trujillo01}. For the HST GSS images, simulations with synthetic data showed that the code provides reliable bulge structural parameters only for galaxies brighter than $F814W\approx 20$ mag, this is a limit too bright for our study. 
Our second and third methods are based on the \GimTwoD\ (Galaxy IMage 2D) bulge-disk decompositions from \citet{Simard02}, available from the DEEP public database.  We defined samples using bulge-to-total ratios ($B/T > 0.4$), or bulge apparent magnitudes ($F814W < 23.57$); the latter criterion was used by \citet{Koo05}.  
Our fourth method was based on concentration and asymmetry indices.  Early-type disk galaxies should have moderately high concentrations and low asymmetries \citep{BershadyConselice2000}.  

Our fifth method was the standard 'mark the disk' method of assuming an exponential profile for the disk, fitting it over some visually-chosen radial range, and selecting galaxies with central surface brightness excess over the inward extrapolation of the outer disk profile.  
We derived averaged profiles from $15\deg$ wedge-shaped apertures opening on both semi-major axes and fit an exponential law to the outer parts of the profiles. 
We defined the \textit{prominence index} $\eta$ as the difference between the measured central surface brightness and the extrapolated central surface brightness of the disk.  

\begin{figure}[!htb]
\begin{center}
\includegraphics[angle=0,width=0.45\textwidth]{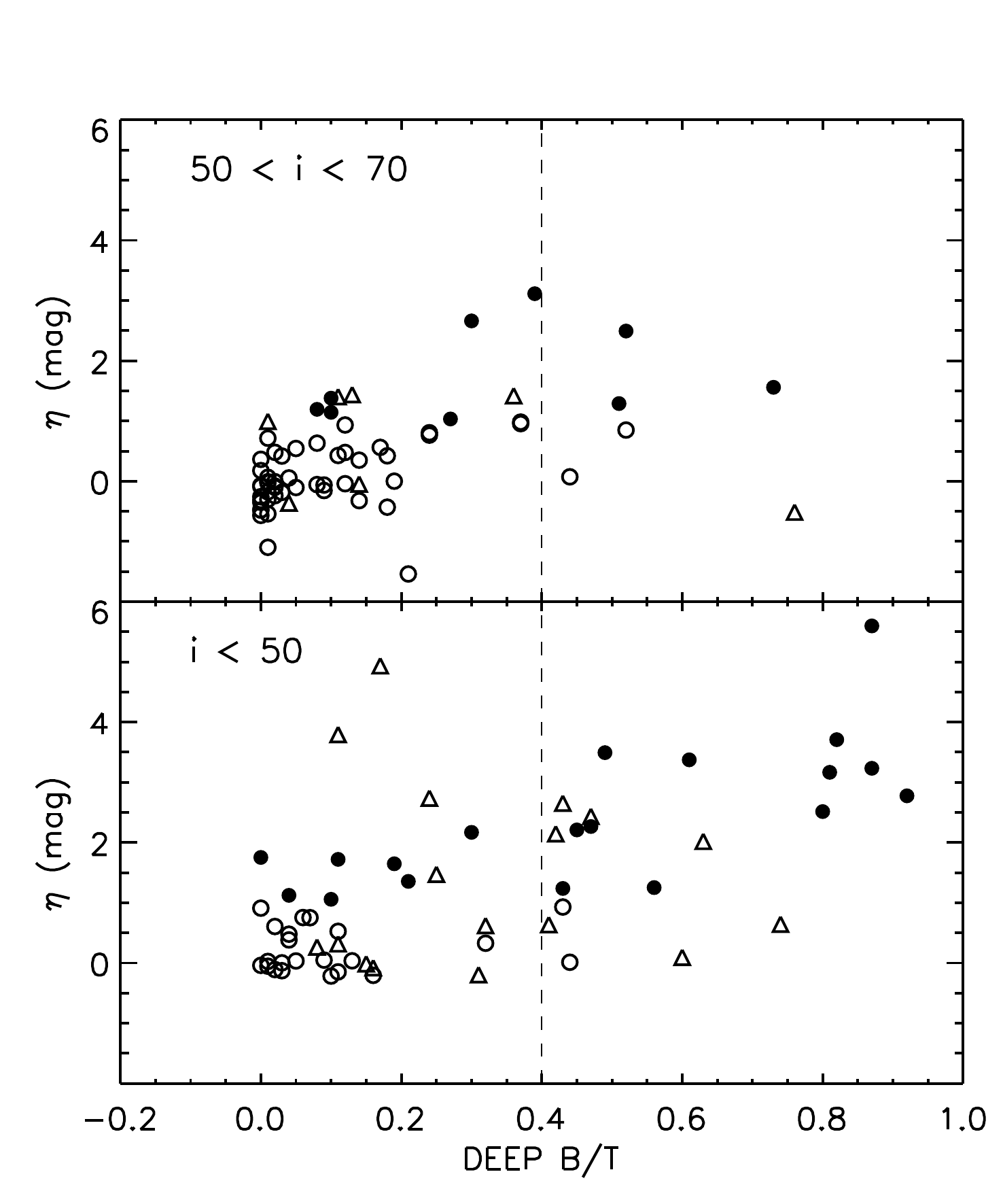}%
\end{center}
\caption{Central prominence index $\eta$ vs DEEP \GimTwoD\ bulge-to-total ratio for the high-inclination and low-inclination samples. {\it Filled circles}: galaxies with $\eta > 1$ (prominent bulges). {\it Open circles}: galaxies with $\eta < 1$. {\it Triangles}: galaxies morphologically classified as mergers.}  \label{fig:BTparam}
\end{figure}

\begin{figure}[!htb]
\begin{center}
\includegraphics[angle=0,width=0.45\textwidth]{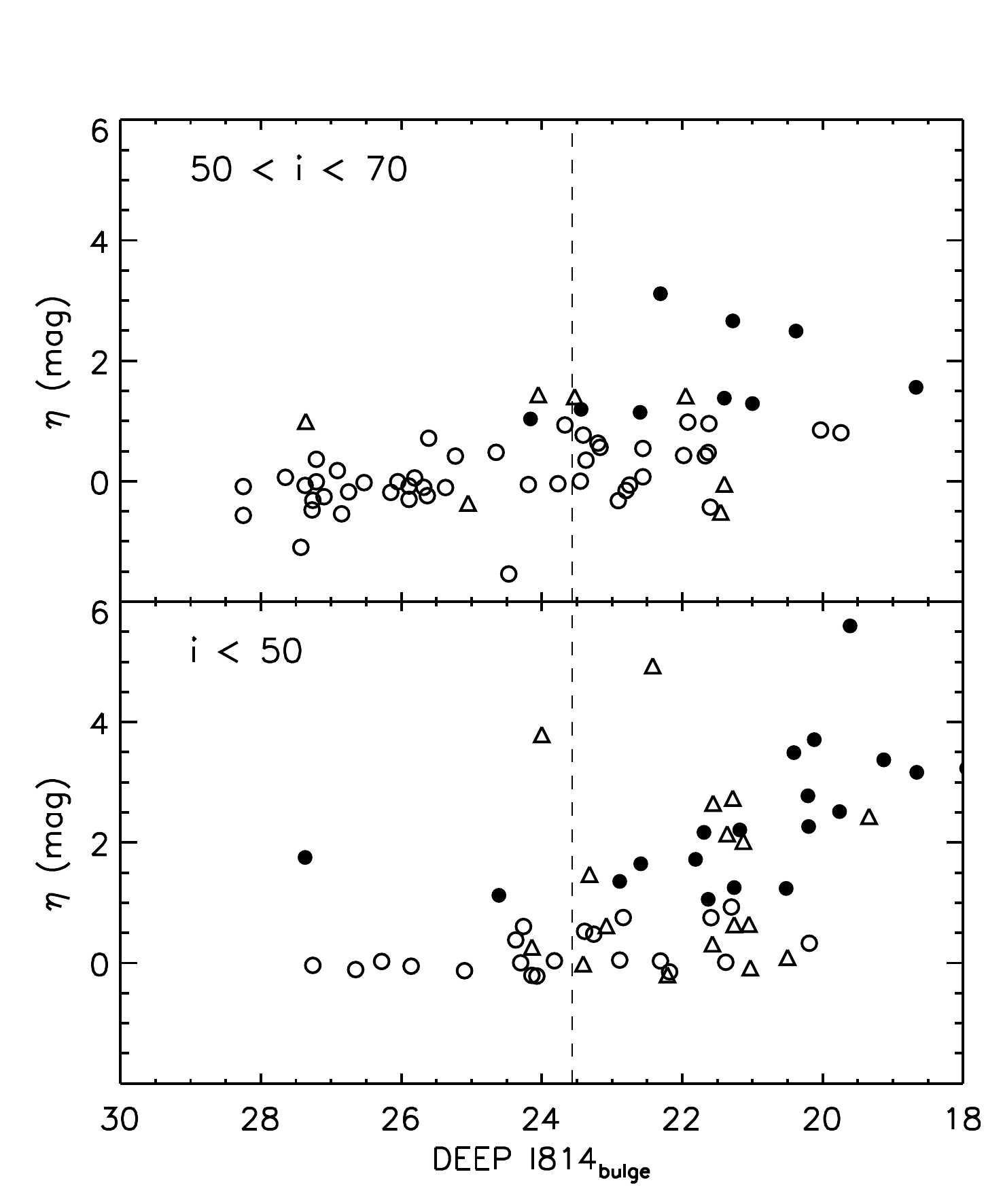}%
\end{center}
\caption{Central prominence index $\eta$ vs DEEP \GimTwoD\ apparent bulge $F814W$ magnitude for the high-inclination and low-inclination samples. {\it Filled circles}: galaxies with $\eta > 1$ (prominent bulges). {\it Open circles}: galaxies with $\eta < 1$. {\it Triangles}: galaxies morphologically classified as mergers.}  \label{fig:Ibulgeparam}
\end{figure}

\begin{figure}[!htb]
\begin{center}
\includegraphics[angle=0,width=0.45\textwidth]{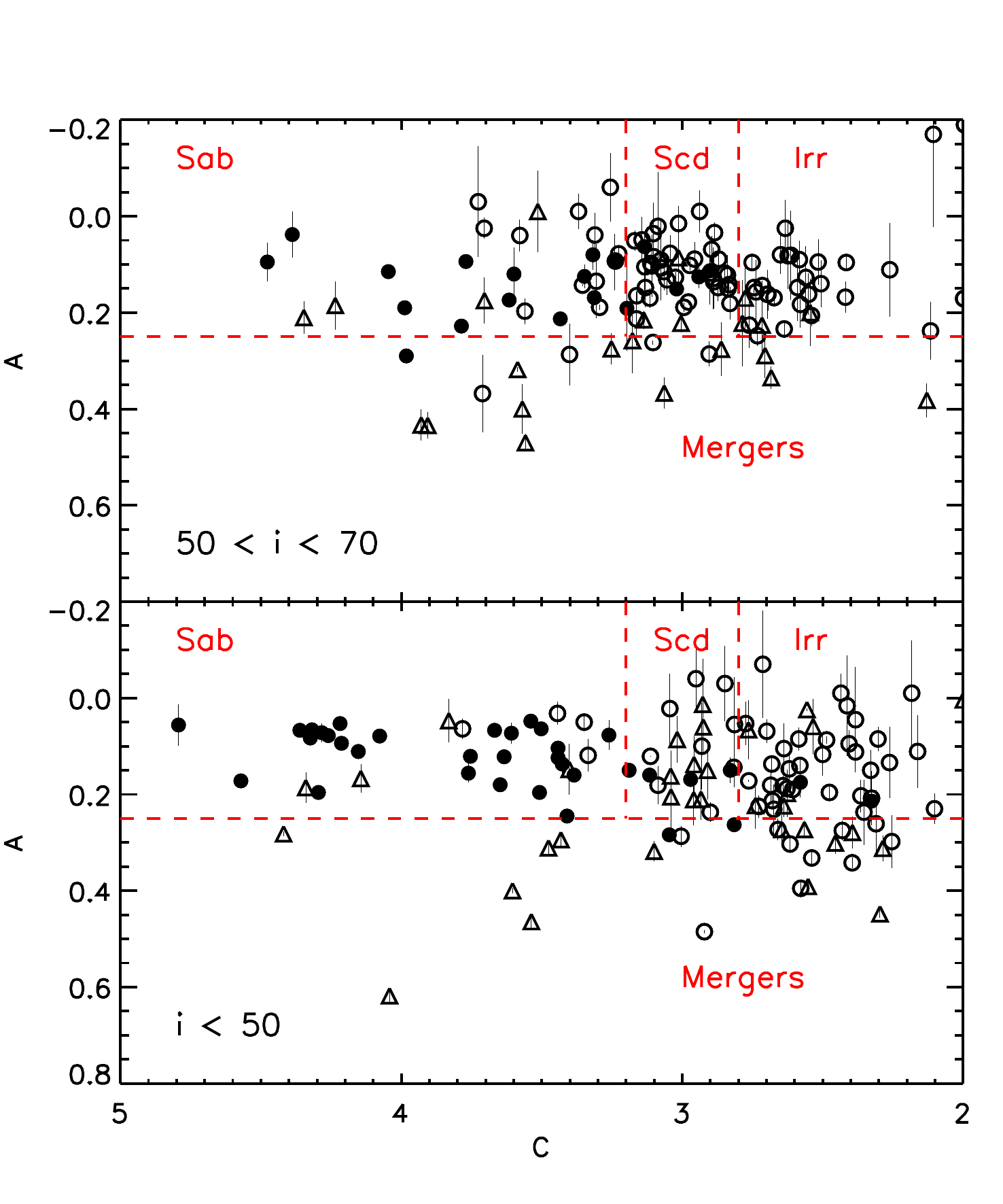}%
\end{center}
\caption{Asymmetry-concentration diagram for the high-inclination and low-inclination samples. {\it Dashed lines} separate different A-C regions populated by different types of galaxies: E/S0/Sab, Scd, Irr, Mergers. {\it Filled circles}: galaxies with $\eta > 1$ (prominent bulges). {\it Open circles}: galaxies with $\eta < 1$. {\it Triangles}: galaxies morphologically classified as mergers.}  \label{fig:CASparam}
\end{figure}

\begin{figure}[!htb]
\begin{center}
\includegraphics[angle=0,width=0.5\textwidth]{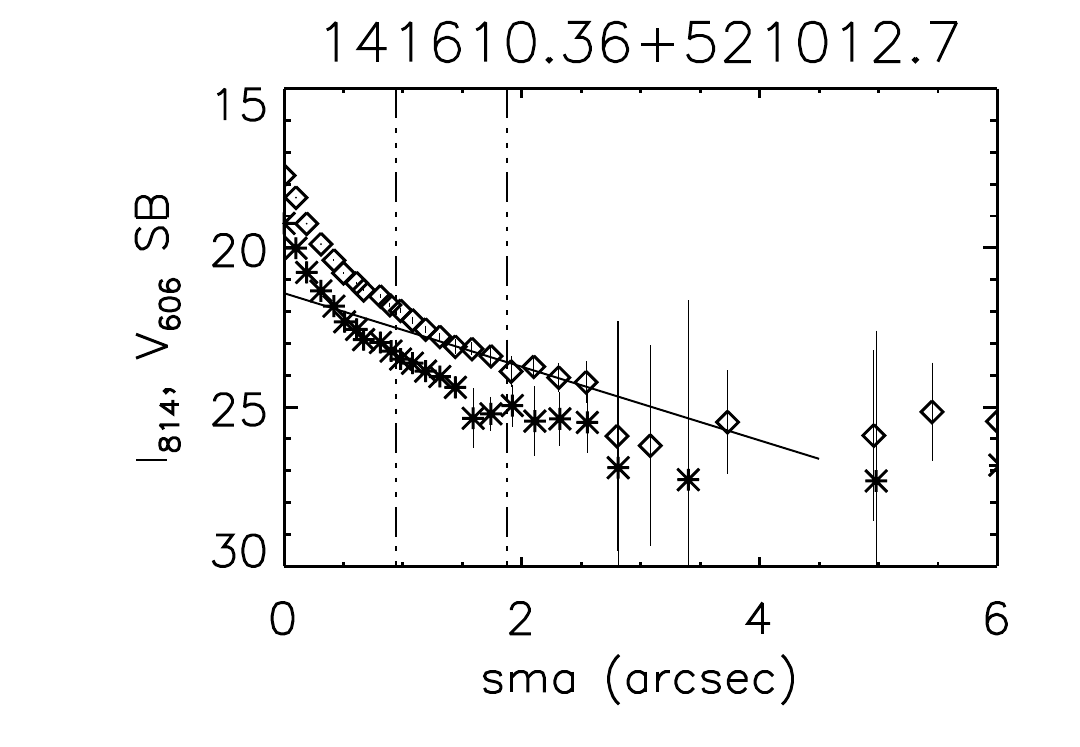}%
\hspace{1in}%
\includegraphics[angle=0,width=0.5\textwidth]{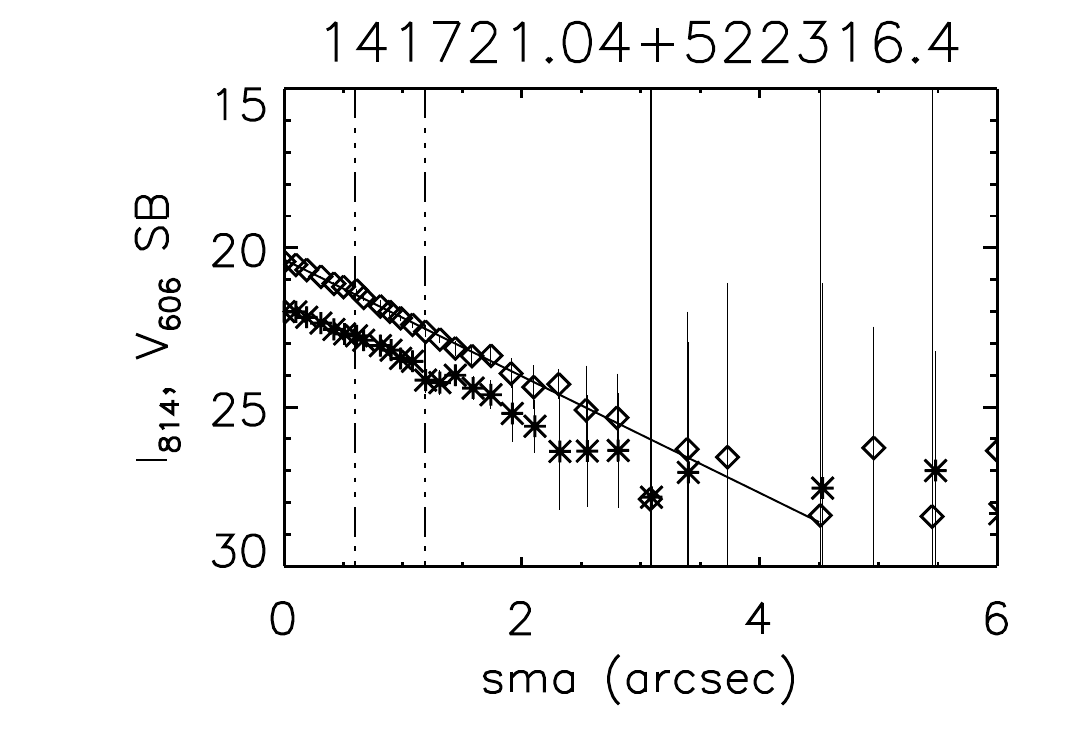}%
\end{center}
\caption{Observed surface brightness averaged profiles derived from 15\deg\ wedge-shaped apertures over both semi-major axes. {\it Diamonds}: $F814W$ band. {\it Asterisks}: $F606W$ band. Both galaxies belong to the high-inclination sample. The {\it solid line} is the exponential law fit over the outer regions of the profiles. {\it Vertical dashed lines}: 1 and 2 times the scale length of the exponential law, respectively; range in which disk colours were measured. The upper galaxy has $\eta >1$, while the lower one has $\eta < 1$.}  \label{fig:selprofiles}
\end{figure}

After inspecting the surface brightness profiles of the sample galaxies, and analysing of how the properties of the bulge sample change with the $\eta$ cutoff, we set a threshold of $\eta = 1$ mag to divide the sample into subsamples \textit{with} and \textit{without measurable bulges}. This was done both for the high- and low-inclination samples. In Fig.~\ref{fig:selprofiles} we show two examples of galaxies with and without central excess belonging to the high-inclination sample.

Before segregating between bulge and non-bulge galaxies, we visually identified objects that could be morphologically classified as mergers, in both high- and low-inclination samples. 
Therefore, we have three subsamples in both low- and high-inclination samples.

The main difficulty with $\eta$ criterion for bulge selection is its level of subjectivity when drawing the exponential fit to the outer profile.  On the other hand, methods based on model fitting such as \GimTwoD\ impose model restrictions whose effects on the model uncertainties are also difficult to quantify.  We argue that the $\eta$ method is simple, involves little modelling and can be verified individually through visual inspection of the profile whenever specific features need to be analysed.

We compare the various bulge sample selection methods outlined above in Figs.~\ref{fig:BTparam}--\ref{fig:CASparam}. 
In Fig.~\ref{fig:BTparam}, we show our prominence index $\eta$ versus $B/T$ from \GimTwoD\ for those galaxies in our samples with a bulge-disk model in the DEEP database.  As expected, most of the $\eta < 1$ galaxies show  $B/T < 0.4$, a $B/T$ range usually ascribed to disk-dominated galaxies.  However,  the $\eta \geq 1$ galaxies have $B/T$ values ranging from $0$ to $1$. Hence, using our $\eta$ criterion, we are able to select smaller-$B/T$ galaxies than when a strict $B/T$ cut is applied.  

In Fig.~\ref{fig:Ibulgeparam}, we compare the $\eta$ selection with that based on bulge apparent $F814W$ magnitudes from the DEEP database. The limit value that \citet{Koo05} used to select their bulge sample was $F814W=23.57$ in the Vega system. We see that most of our $\eta > 1$ galaxies have a DEEP bulge apparent $F814W$ magnitude brighter than the \citet{Koo05} cutoff, so both criteria coincide in this aspect, but there are also galaxies with $\eta < 1$ and mergers for which DEEP lists a bright bulge apparent magnitude.  We interpret this as an indication that selecting by DEEP bulge magnitudes yields a sample that includes galaxies without bulge-disk structure, which we would classify in the non-bulge disk sample.  

The distribution of the $\eta$-selected samples in the concentration-asymmetry plane is illustrated in Fig.~\ref{fig:CASparam}. The region of low asymmetry and high concentration should be populated by early-type galaxies: E/S0/Sa/Sab. In the low-inclination sample, the concentration-asymmetry does select galaxies with $\eta > 1$.  But for high-inclination galaxies the method is of limited use as the concentration index does not discriminate well between galaxies with and without prominent bulges because of the high surface brightness of inclined disks.  For low-inclination galaxies, the method has the additional problem of including bonafide ellipticals with the bulge sample, something we want to avoid, as discussed in Sect.~\ref{sec:approach}.

\subsubsection{Sample contamination by elliptical galaxies}
\label{sec:ellipticalcontamination}

After segregating between the three types of subsamples (bulge galaxies, non-bulge galaxies and mergers) we visually inspected the bulge sample to estimate the pollution of the sample by elliptical galaxies. While few ellipticals are expected in the high-inclination sample (the ellipticity cut allows only E7 galaxies), ellipticals may be present in the low-inclination sample.  
The criteria we used to identify them were the morphology as well as the surface brightness profiles and the photometric SEDs. We looked for galaxies with smooth structure, lack of evidence for spiral arms or localised star forming regions, with evidence of one single sersic surface brightness profile (no exponential outer component) and with very red SED typical of old populations.
We find 5 possible elliptical candidates from 54 ($9.3\%$ of the bulge sample); all but one belong to the low-inclnation sample. The id number of those objects are: 1138, 1137, 1113, 160, 1065.

\section{Derivation of colours}
\label{sec:methodology}

\subsection{Wedge colour profiles}
\label{sec:profiles}

For the derivation of bulge colours, we worked with colour profiles along wedge-shaped apertures opening on the minor axis.  We chose this method, over the traditional one of using elliptically-averaged surface brightness profiles, to reduce the amount of reddening from dust in the disk, which can be quite important in inclined galaxies. A comparison of colour profiles along both semi-minor axes with wedges indicates which of the two bulge sides is reddened by the disk; this reddened side is discarded, and the bulge colour is extracted from the other profile. While we lose some signal with this method, for inclined galaxies the derived profiles gain reliability as dust reddening effects are greatly diminished. Even at intermediate redshifts, where the wedge method has some limitations (see Sect.~\ref{sec:approach}), it is preferred to the ellipse method. For low-inclination galaxies, the differences are small. 

\begin{figure}[!htb]
\begin{center}
\includegraphics[angle=0,width=0.2\textwidth]{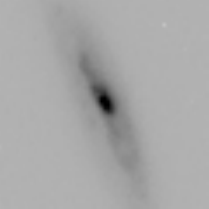}%
\includegraphics[angle=0,width=0.2\textwidth]{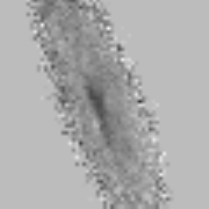}%
\hspace{2in}%
\includegraphics[angle=0,width=0.45\textwidth]{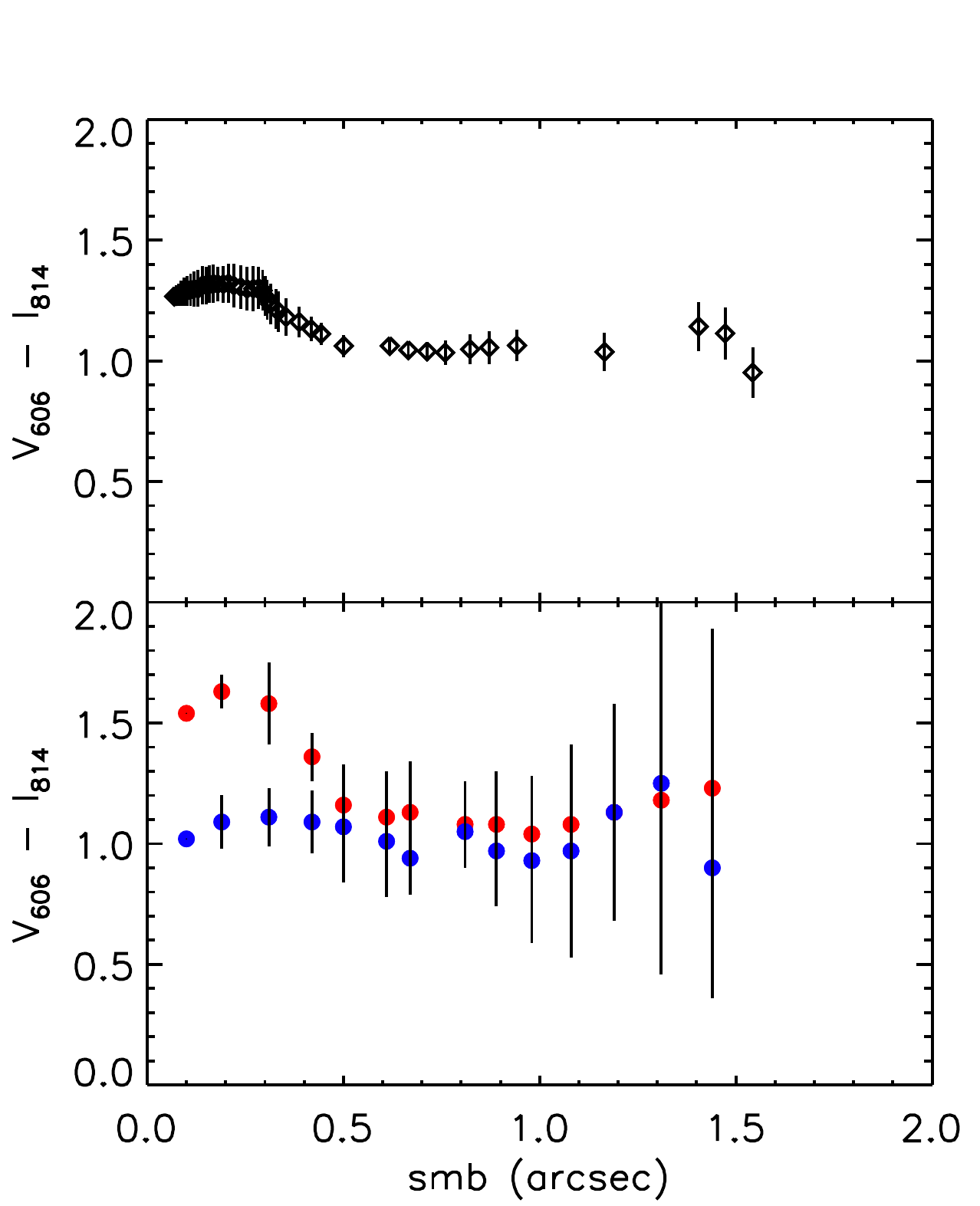}%
\end{center}
\caption{
(\textit{a}) Surface brightness map and colour map for an inclined galaxy. The map side is 6\arcsec. Darker areas correspond to brighter and redder regions, respectively. 
(\textit{b})
Observed $(F606W-F814W)$ colour profile derived from a elliptically-averaged surface brightness profile. 
(\textit{c}) Colour profiles derived along 40\deg\ wedge-shaped apertures opening on both semi-minor axes; {\it red points:} side reddened by dust in the disk; {\it blue points:} profile that diminishes dust reddening effect since its light is not seen through the disk.}  \label{fig:comparationmet}
\end{figure}

Figure~\ref{fig:comparationmet} shows an example of colour profiles of an inclined galaxy derived using wedges and using elliptically-averaged surface brightness profiles. The example shown ($\epsilon=0.75$, disk inclination $i=75\deg$) is slightly above our high-inclination limit, and hence does not belong to our final sample; however, it provides a clear, extreme example of the effects of dust on bulge colours.  
In Fig.~\ref{fig:comparationmet}a, we show the $F814W$-band image and the $(F606W-F814W)$ colour map in gray scale. Red linear bands along the galaxy disk to the left of the centre indicate that the disk lies in front of the bulge on the left side of the image.
An elliptically-averaged colour profile, shown in Fig.~\ref{fig:comparationmet}b, shows a central region of red colours that would suggest the bulge region is distinctively redder than the disk; however, the minor-axis colour profiles (Fig.~\ref{fig:comparationmet}c) show that only one minor axis is redder than the disk, namely, the one to the left of the centre, in which the bulge is seen through the disk.  Dust reddening is responsible for the redder colours; by measuring bulge colours on the side of the bulge that lies above the disk (to the right of the centre in Fig.~\ref{fig:comparationmet}a), dust reddening is largely avoided.  
Note that, in the case where the intrinsic colour structure of the galaxies consisted of (a) a red bulge surrounded by a disk with bluer populations, and (b) an opaque dust lane in the mid plane, with scale height much smaller than that of the stellar disk \citep[a configuration studied by][]{Tuffs04,Moellenhoff06}, then dust would strongly attenuate half the galaxy light. However, it would only moderately redden the light that gets emitted on the near side of the galaxy.  The result would be a colour asymmetry on the two semi-minor axes, but of opposite sign to the one due to dust described above, namely, the red side would correspond to the bulge.  
Such configuration does not seem to apply to hardly any of our galaxies, given that features in the colour maps do not follow the elliptical isophotes of the bulges, but tend to define linear patches corresponding to dust lanes (see Fig.~\ref{fig:comparationmet}a).  The same was found at $z=0$ in a similarly selected disk galaxy sample by \citet{Balcells94}: the redder semi-minor axis corresponds to the axis with the disk in front of the bulge.

The direction of the major axis was that given by \SExtractor. Wedges on each semi-minor axis had an opening angle of $45\deg$.  The vertex of the wedges were placed at the centre of the $F814W$ galaxy image, as determined with the IRAF task \textsc{imcenter}.  
The $F814W$ and $F606W$ images are registered to within $0.24$ pixels, and we did not correct for such small offcentring.  
The local background level was estimated with \SExtractor\, using a background mesh width of 96 pixels and a local annulus thickness of 24 pixels.

Disk colours were similarly extracted from wedge apertures opening onto the major axis and with an aperture angle of 15$\deg$. Again, for inclined galaxies with dust in the disk, wedge apertures are preferred over the classical method of combining elliptically-averaged surface brightness profiles. 

\subsection{Nuclear colours}
\label{sec:bulgecolour}

Throughout the paper, we will use local colours measured at fixed physical distances of $0.85$ kpc from the centre, on the bluer semi-minor axis, 
as estimates of bulge colours. This distance is reasonably below the typical size of bulges, both in the local Universe \citep[see][for bulge-disk decompositions for the Millenium catalog]{Allen06} and at intermediate redshift in the Hubble Deep Field (HDF) \citep{Trujillo04}. The former finds that the main bulk of galaxies have bulges with $0.9< \Reff <4$ kpc, while \citet{Trujillo04} get a median \Reff\ of their bulges equal to 1.07 kpc. In addition, by measuring the colour at this galactocentric distance, we minimise the effects of any residual reddening due to dust. The measured colour is close to the intrinsic colour of the stellar populations, since at 0.85 kpc from the centre, the stellar population scale height is still much larger than the one of the dust layer on the disk. We also measure the colour at the same galactocentric distance in the non-bulge sample, in order to compare the bulge and the non-bulge sample colours.

The physical distance $0.85$ kpc corresponds to an angular size of $\sim 0.3$\arcsec\ and $\sim 0.1$\arcsec\ at $z=0.15$ and $z=1.2$, respectively. This radius lies very close to the centre for $z > 0.9$ and nuclear colours may be biassed by the image PSF, however measurement of the PSF in the $F814W$ and $F606W$ WFPC2 GSS images revealed a median difference in PSF width of $\sim 0.1 $ pixels.  Due to the poor sampling of WFPC2 PSF ($\sim 1.4$ pix per FWHM), we decided not to correct the PSF difference between $F814W$ and $F606W$.  We do not expect derived colours to be affected by such small PSF differences.  

As already explained in Sect.~\ref{sec:approach}, we do not apply any correction from disk light.  Our results support this choice, as the colour structure of our galaxies does not correspond to a redder bulge surrounded by a uniformly bluer disk (see details of the analysis of colour profiles in Paper II, Sect. 4). 

\begin{figure*}[!htb]
\begin{center}
\includegraphics[angle=0,width=0.90\textwidth]{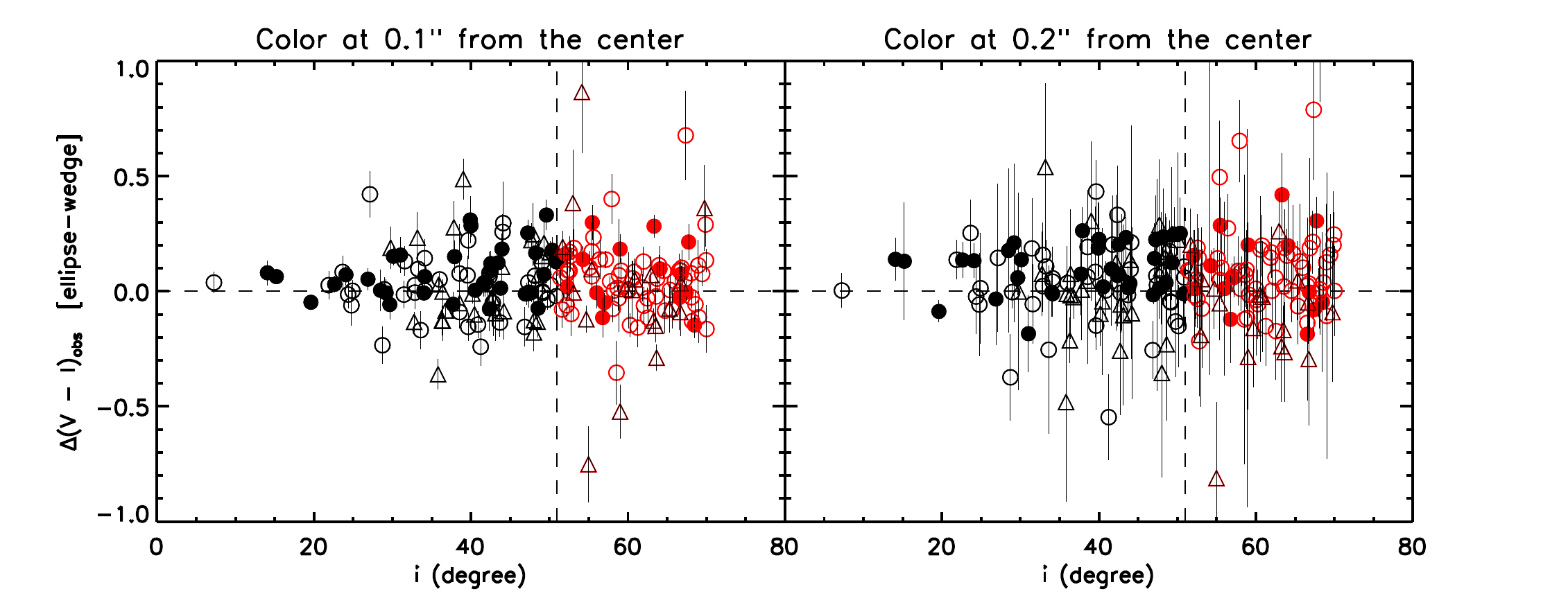}%
\end{center}
\caption{Differences between colours calculated by ellipse method and the colours by wedge method at different distances from the centre ({\it left}: 0.1'', {\it right}: 0.2'') vs the galaxy inclination. {\it Filled circles}: colours of galaxies with $\eta > 1$, prominent bulges. {\it Open circles}: colours of galaxies with $\eta < 1$. {\it Triangles}: galaxies morphologically classified as mergers. The {\it vertical dashed line} separates the low- and high-inclination samples ({\it black} and {\it red points}, respectively, in the electronic format.}  \label{fig:vi_comp_deep_ellip}
\end{figure*}

Globally, the wedge colours are slightly bluer than those derived from ellipse-fitting colour profiles. We show the colour differences (ellipse \textit{minus} wedge) vs the galaxy inclination in Fig.~\ref{fig:vi_comp_deep_ellip}.  The scatter increases toward higher inclinations. We have calculated the median values of colour differences in galaxies with high central excess ($\eta > 1$) for three inclination bins, $i < 30$\deg, $30$\deg\ $ < i < 55$\deg\, $55$\deg\, $<i<70$\deg. We obtain for a distance from the centre equal to $0.1$\arcsec\, 0.05, 0.12 and 0.08, with standard deviations 0.07, 0.12, 0.74,  respectively. At $0.2$\arcsec\ the median values and the standard deviations for the same inclination bins are, respectively, 0.13, 0.11, 0.06 and 0.12, 0.10, 0.57. 

\subsection{Disk colours and aperture colours}
\label{sec:diskcolours}

We estimate disk colours by averaging in the range from 1 to 2 times the scale length of the exponential law fit to the outer regions of the galaxies in the average of the profiles derived from wedge-shaped apertures over both semi-major axes. 
Contrary to bulges, which we expect to have bigger scale heights above the galaxy mid-plane, stars in disks are expected to be mixed with dust, and hence, the measured disk colours may be affected by dust reddening.

To obtain an estimate of the global galaxy colours, we use colours extracted from fixed circular apertures of diameter 2.6\arcsec, on images that are smoothed to a common FWHM of 1.3\arcsec.  Such apertures fail to encompass the entire galaxies in some cases, but they provide representative global colours given the little light in the outer parts and the fact that colour gradients are very gentle.  The comparison of global colours and disk colours with bulge colours is analysed in Paper II.  

\subsection{Colour transformations and absolute magnitude computation}
\label{sec:colourtransformation}

We calculated K-corrections using SEDs covering bands $U, B, F606W, F814W, J$, and \Ks\ from the GOYA photometric survey. Templates were obtained from the best-fit solutions delivered by \Hyperz, which was allowed to explore solutions over a narrow $z$ range around the redshift of the galaxy.  

Our approach differs from that of \citet{Koo05}, who convert observed ($F606W-F814W$) colours into rest-frame ($U-B$) with parametric conversions derived from a subset (34 spectra) of an atlas of 43 spectra of local galaxies \citep{Kinney96}. The use of theoretical SEDs from stellar population synthesis has potential disadvantages over the use of empirical models.  Synthetic templates do not include factors that affect the SED, such as variation in metallicity and emision lines. Moreover, we are applying K-corrections, calculated from integrated SEDs of the whole galaxy, to both the bulge and the disk. However, for the galaxies in common with the sample of \citet{Koo05}, a comparison of the K-corrections shows that differences are indeed small, with a standard deviation $\sim 0.05$, which is dominated by our three reddest galaxies that are $\Delta (U-B)=0.1$ mag redder than if we use Koo's K-corrections. The error in K-corrections derived from photometric redshift uncertainty and from SED choice have median values $\Delta K_{606} \sim 0.03$ and $\Delta K_{814} \sim 0.03$. 

The $(U-B)$ and $(B-R)$ colour indices were chosen to provide rest-frame colour data. The index $(U - B)$ roughly coincides with observed $(F606W - F814W)$ at redshifts around $z \thicksim 0.8$, while $(B-R)$ matches the observed filters at redshifts around $z \thicksim 0.3$, near the low-end of our redshift range.  With these choices, K-correction errors cancel out with SED colour errors.  Additionally, much data is available on $(B-R)$ colours for galaxies in the local Universe, making the $(B-R)$ colour particularly useful.  Therefore, throughout Paper II, all important relations will be shown in both $(U-B)$ and $(B-R)$.  We will see that colour distributions in both sets of colours describe very similar patterns.  

The colour transformations were done using the relations~\ref{eqn:filtroUB} and ~\ref{eqn:filtroBR}:

\begin{eqnarray}
(U - B)^{\rm o} = (F606W - F814W)^{\rm obs} - {\rm K}_{F606W} + {\rm K}_{F814W} \nonumber \\
+ (U - F606W)_{\rm SED}^{\rm o} - (B - F814W)_{\rm SED}^{\rm o}
\label{eqn:filtroUB}
\end{eqnarray}

\begin{eqnarray}
(B - R)^{\rm o} = (F606W - F814W)^{\rm obs} - {\rm K}_{F606W} + {\rm K}_{F814W} \nonumber \\
+ (B - F606W)_{\rm SED}^{\rm o} - (R - F814W)_{\rm SED}^{\rm o} 
\label{eqn:filtroBR}
\end{eqnarray}

\noindent where the superindices $^{\rm o}$ indicate that the parameter is in the rest frame of the galaxy, $(F606W - F814W)^{\rm obs}$ is the colour in the observer's frame and $(U - F606W)_{\rm SED}^{\rm o}$, $(B - F814W)_{\rm SED}^{\rm o}$, $(B - F606W)_{\rm SED}^{\rm o}$ and $(R - F814W)_{\rm SED}^{\rm o}$ are the $(U - F606W)$, $(B - F814W)$, $(B - F606W)$ and $(R - F814W)$ colours in the rest frame, calculated by using the theoretical SED of the galaxy.

The error in $(U-B)$ and $(B-R)$  derived from K-correction errors and redshift uncertainty have median values $\Delta (U-B) \sim 0.15$ and $\Delta (B-R) \sim 0.12$. Those errors must be added in quadrature to the measured errors of the observed colours tabulated in Tables~\ref{tab:bul_colours} and~\ref{tab:nobul_colours}.

The absolute magnitudes are given by

\begin{equation}
M_i = m_i + 5 - 5 \log{D_{\rm L}} - {\rm K}_i
\label{eqn:absmag}
\end{equation}

\noindent where $M_i$ is absolute magnitude in $i$-band, $m_i$ is apparent magnitude also in $i$ band, $D_{\rm L}$ is the luminosity distance, and K$_i$ is the K-correction in $i$ band. The error in absolute magnitudes derived from redshift uncertainty and from SED choice have median values $\Delta M_B \sim 0.70$ and $\Delta M_R \sim 0.57$. This error must be added in quadrature to the measured errors of the observed magnitude $m_i$. 

\section{Results}
\label{sec:results}

We summarize the results of this paper in the Tables~\ref{tab:bul_datos}, \ref{tab:nobul_datos}, \ref{tab:bul_colours}, \ref{tab:nobul_colours}, \ref{tab:bul_gradient}, and \ref{tab:nobul_gradient} in Appendix~\ref{sec:datatables}. Tables~\ref{tab:bul_datos} and \ref{tab:nobul_datos} give basic source parameters for galaxies in the bulge sample and pure disk sample, respectively.  We include the source ID, the redshift and redshift quality (Sect.~\ref{sec:redshifts}), the galaxy semi-major axis and ellipticity (Sect.~\ref{sec:disksample}), the total apparent $F814W$ magnitude, and absolute magnitudes in Johnson-Cousins $B$ and $R$.  In these and the rest of the tables in Appendix~\ref{sec:datatables}, galaxies are sorted in order of increasing redshift. Tables~\ref{tab:bul_colours} and \ref{tab:nobul_colours} list observed $(F606W - F814W)$ colours (Sects.~\ref{sec:bulgecolour},~\ref{sec:diskcolours}), as well as inferred rest-frame $(B-R)$, and $(U-B)$ colours (Sect.~\ref{sec:colourtransformation}), for bulge (nucleus), disk, and galaxy for the samples with and without bulges, respectively. Finally, Tables~\ref{tab:bul_gradient} and \ref{tab:nobul_gradient} show the colour gradients for each galaxy in both semi-minor axes colour profiles and also in the averaged semi-major axis one.

Postage stamps, colour maps, spectral energy distributions, surface brightness, and colour profiles are provided for all of the sources in Appendix~\ref{sec:samplesbulges}. Colour profiles over the geometrically deprojected bluer semi-minor axes in kpc are shown Appendix~\ref{sec:bluercolprof}.

The numbers of objects in each of our defined subsamples (see Sect.~\ref{sec:sample})
are given in Table~\ref{tab:SampleNumbers}.  The total sample is composed of 312 objects, of
which 42\% have spectroscopic redshifts and 37.5\% have photometric
redshifts.  The main subsample of objects with known redshifts thus
comprises 248 galaxies.  Of these, 54 are disk galaxies with measurable
bulges (21.8\% of the sample with known redshift), 137 are bulgeless disk galaxies (55.2\%), and
57 are mergers (23\%). Four low-inclination galaxies and one high-inclination galaxy from the bulge sample (9.3\%) are possible elliptical candidates. All subsamples include a number of dwarf galaxies, defined as having $M_B > -18$; the dwarf frequency is higher in the high-inclination subsample due to the higher surface brightness of inclined disks.

\begin{table}[htdp]
\caption{Number of galaxies in each subsample}
\begin{center}
\begin{tabular}{lcc}
\hline
\hline
Sample & \multicolumn{1}{c}{$i<50$}	& \multicolumn{1}{c}{$50< i < 70$}	\\ 
\hline
$R > 1.4\arcsec$	& 142		& 	170			\\
\zspec		& 63		& 68			\\
\zphot		& 62		& 55			\\
$\eta > 1$	 (bulges)		&	35		&	19			\\
Ellipticals  & 4 & 1 \\
Mergers				&	34		&	23			\\
$\eta < 1$	 (no bulges)	&	56		&	81			\\
$M_B < -18$  &  5  & 14  \\
\hline
\hline
\label{tab:SampleNumbers}
\end{tabular}
\end{center}
\end{table}%

\begin{figure}
\begin{center}
\includegraphics[angle=0,width=0.45\textwidth]{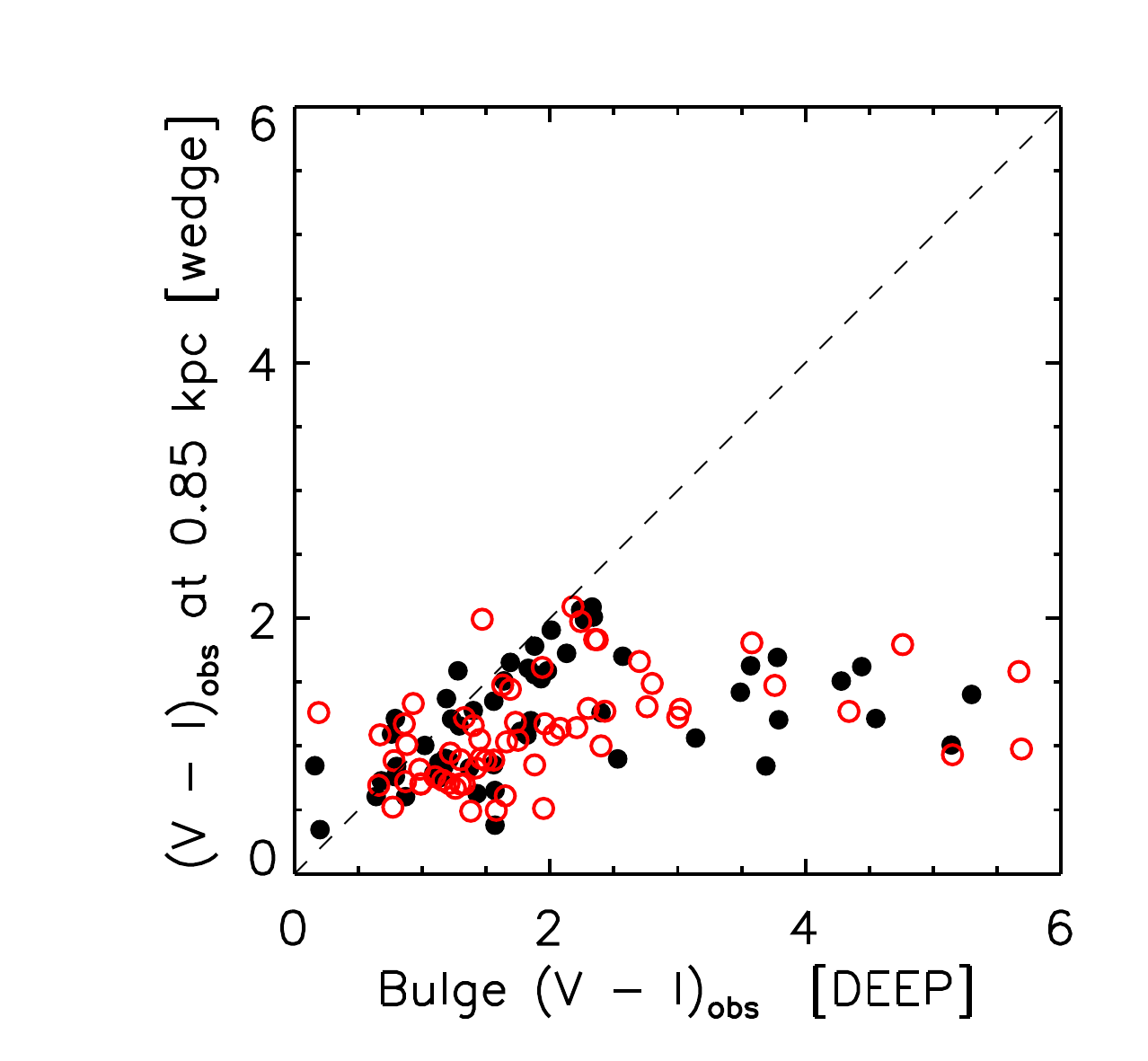}%
\end{center}
\caption{Observed $(F606W-F814W)$ wedge colour measured at 0.85 kpc vs integrated bulge colour from DEEP bulge-disk decomposition, for the 123 galaxies in our sample for which the DEEP database lists a bulge colour. {\it Filled circles}: galaxies from the low-inclination sample. {\it Open circles}: galaxies from the high-inclination sample.}  \label{fig:vi_comp_deep}
\end{figure}

\subsection{External comparison}
\label{sec:extcomparison}

In Fig.~\ref{fig:vi_comp_deep}, we show the differences between wedge
nuclear colours and \GimTwoD\ bulge colours, available from the DEEP
database, for the 123 galaxies in our total sample for which the DEEP
database lists a bulge colour. For most galaxies, \GimTwoD\ colours are
redder than our wedge colours.  The main cloud of the distribution
scatters within $0 < \Delta(F606W-F814W) < 1$, for those we get a median difference value equal to
0.28 and a standard deviation equal to 0.35. 
Such offset is significant.  
When converted to rest-frame $(U-B)$
colours, it is larger than the $\Delta(U-B) = 0.17$ reddening expected for passive
evolution from $z=1$ to $z=0$ (assuming here $\zform = 3.0$; see Paper II, Sect. 7, Fig. 2).  

Additionally, $30\%$ of the galaxies have DEEP bulge colours dramatically redder than our colours ($\Delta(F606W - F814W) = 1-5$). Those are objects whose DEEP bulge colours have very large errors, $>1.2$ mag, and we suspect that the bulge-disk decomposition is particularly unstable in those cases.  

 Even after excluding the highly-deviant cases, the colour differences between DEEP and us highlight the difficulties
in assigning a single, representative colour for bulges of intermediate
redshift galaxies. Part of the offset must come from the fact that, in
the derivation of bulge colours in the DEEP database, disk light gets
subtracted assuming a uniform colour for each component. Because, in
fact, most galaxies show a negative colour gradient (bluer outward, see colour profiles in Appendix~\ref{sec:samplesbulges}), when
assuming a uniform disk colour, one subtracts light that is too blue from the 
bulge region; this yields a bulge colour biassed to the red. We made
simple estimates of disk contamination under the assumption of an
exponential disk with uniform colour. Colour corrections range from
$\Delta(F606W-F814W) \sim 0.0$ for the highest $\eta$, up to $\Delta(F606W-F814W) \sim
0.5$ for $\eta \sim 1$: the mean bulge colour correction for the high-
(low)-inclination subsamples is 0.13 (0.03). 

Given that these corrections are smaller than the median colour difference
between the DEEP measurement and ours, other processes may be at work. Plausibly, forcing a
\RdeV\ funtional form for the bulge profile, and an exponential profile
for the disk, might contribute to making the bulge redder as well. 
Conversely, our bulge colours may be biassed to the blue if our wedge
apertures happen to hit nuclear patches of star formation or, in cases in which the bulge side does not correspond to the bluest semi-minor axis (see Sect.~\ref{sec:profiles}).

\section{Conclusions}
\label{sec:conclusions}

We have presented a sample of 248 disk galaxies from the Groth strip with known redshifts in which we have identified candidate bulge components. We provide global photometry as well as  colour measurements for bulges, disks, and for each galaxy as a whole.  For galaxies without bulges, we provide colour measurements at the same radial distances and following the same method as done for the galaxies with bulges.  These colours are analysed in Paper~II.  The tables in the Appendices are available in electronic format from the authors.  

\begin{acknowledgements}
We thank the anonymous referee for suggestions that improved the paper and Ignacio Trujillo, Carlos L\'opez, David Abreu, Marc Vallb\'e, Enrique Garc\'\i a-Dab\'o, Alfonso Arag\'on-Salamanca, David Koo, and Reynier Peletier, for useful discussions. 
This work was supported by the Spanish Programa Nacional de Astronom\'\i a y Astrof\'\i sica through project number AYA2006-12955. 
Some/all of the data presented in this paper were obtained from the Multimission Archive at the Space Telescope Science Institute (MAST). STScI is operated by the Association of Universities for Research in Astronomy, Inc., under NASA contract NAS5-26555. Support for MAST for non-HST data is provided by the NASA Office of Space Science via grant NAG5-7584 and by other grants and contracts.
This work uses data obtained with support of the National Science Foundation grants AST 95-29028 and AST 00-71198.
\end{acknowledgements}

\bibliographystyle{aa}
\bibliography{9406refs.bib}

\clearpage

\onecolumn

\begin{appendix}
\section{Data tables}
\label{sec:datatables}

\noindent Notes for Tables~\ref{tab:bul_datos} and~\ref{tab:nobul_datos}:

\noindent (1) Source number identification \\
(2) Source ID, given by sky coordinates of the source: RA+DEC  \\
(3) Source redshift   \\
(4) Redshift quality: 1 - spectroscopic, 2 - photometric  \\
(5) Source semi-major axis in arcsec from \SExtractor\ "A" size parameter   \\
(6) Source ellipticity from \SExtractor\ $(1-b/a)$   \\
(7) Subsample from which the source comes: hi - high-inclination sample, li - low-inclination sample   \\
(8) Total $F814W$ magnitude of galaxy from \SExtractor\ $MAG\_BEST$ magnitude)  \\
(9) Galaxy absolute magnitude in $B$-band \\
(10) Galaxy absolute magnitude in $R$-band  \\

\noindent \footnotesize{Note for columns 9--10: The total errors in galaxy absolute magnitude, $\Delta M_B$ and $\Delta M_R$, come
from the sum in quadrature of the measured error in observed $F814W$ magnitude and the error derived from redshift uncertainty and SED choice (see
Sect.~\ref{sec:colourtransformation}).} \\

\noindent Notes for Tables~\ref{tab:bul_colours} and~\ref{tab:nobul_colours}:

\noindent (1) Source number identification \\
(2) Observed nuclear $(F606W-F814W)$ colour, measured at 0.85 kpc from the centre  \\
(3) Observed disk $(F606W-F814W)$ colour, measured in the range between 1 and 2 times the scale length of the exponential law fit to the outer regions of the galaxies  \\
(4) Observed global $(F606W-F814W)$ colour, measured within a 2.6\arcsec\ diameter aperture  \\
(5) Rest-frame nuclear $(U-B)$ colour   \\
(6) Rest-frame disk $(U-B)$ colour   \\
(7) Rest-frame global $(U-B)$ colour \\
(8) Rest-frame nuclear $(B-R)$ colour.   \\
(9) Rest-frame disk $(B-R)$ colour   \\
(10) Rest-frame global $(B-R)$ colour   \\

\noindent \footnotesize{Note for columns 5--10: The total errors in rest-frame colours, $\Delta(U-B)$ and $\Delta(B-R)$, come
from the sum in quadrature of the measured error in $(F606W-F814W)$ colour and the error derived from the colour transformation (see
Sect.~\ref{sec:colourtransformation}).} \\

\normalsize
\noindent Notes for Tables~\ref{tab:bul_gradient} and~\ref{tab:nobul_gradient}:

\noindent (1) Source number identification \\
(2) Colour gradient calculated over the bluer deprojected semi-minor axes colour profile \\
(3) Colour gradient calculated over the redder deprojected semi-minor axes colour profile \\
(4) Colour gradient calculated over the averaged semi-major axes colour profile \\

\noindent \footnotesize{Note for columns 2--4: The colour gradients are measured in
magnitudes per deprojected kpc. The errors are the 1-sigma uncertainty estimates for the returned parameters of the linear
fits.}

\longtab{1}{
\begin{longtable}[!hbp]{r c l c c c c c c c}
\caption{Bulge sample data} \label{tab:bul_datos} \\
\hline
\hline
No. & Source ID  & $z$ &  Qz &  R (arcsec) & $\epsilon$ & Sample & $I_{814}$  & $M_B$ & $M_R$\\
(1) & (2) & (3) & (4) & (5) & (6) & (7) & (8) & (9) & (10)\\
\hline
\endfirsthead
\caption{Continued.} \\
\hline
\hline
No. & Source ID  & $z$ &  Qz &  R (arcsec) & $\epsilon$ & Sample & $I_{814}$  & $M_B$ & $M_R$\\
(1) & (2) & (3) & (4) & (5) & (6) & (7) & (8) & (9) & (10)\\
\hline
\endhead
\hline
\hline
\endfoot
 1131  & 141707.44+522439.8   &      0.045   &      2  &   2.62    & 	 0.28  &  li  &     $18.44 \pm 0.01$ &    -16.89  &     -17.63  \\
 1139  & 141524.68+520554.1   &      0.095   &      2  &   3.84    & 	 0.24  &  li  &     $17.64 \pm 0.01$ &    -18.72  &     -20.04  \\
 1078  & 141611.84+521312.2   &     0.1118   &      1  &   1.78    & 	 0.09  &  li  &     $19.69 \pm 0.01$ &    -17.41  &     -18.42  \\
 1081  & 141646.71+521930.3   &       0.15   &      2  &   1.82    & 	 0.33  &  li  &     $18.67 \pm 0.01$ &    -18.70  &     -20.10  \\
 1125  & 141624.06+521517.0   &     0.1915   &      1  &   2.46    & 	 0.04  &  li  &     $18.81 \pm 0.01$ &    -19.19  &     -20.56  \\
  139  & 141547.15+520610.5   &      0.235   &      2  &   2.26    & 	 0.52  &  hi  &     $20.76 \pm 0.02$ &    -18.11  &     -19.14  \\
  170  & 141724.80+522553.5   &     0.2381   &      1  &   4.17    & 	 0.39  &  hi  &     $19.10 \pm 0.02$ &    -20.03  &     -20.83  \\
  159  & 141718.47+522419.7   &     0.2811   &      1  &   2.95    & 	 0.39  &  hi  &     $18.56 \pm 0.01$ &    -20.41  &     -21.82  \\
\end{longtable}}

\longtab{2}{
\begin{longtable}[!hbp]{r c l c c c c c c c}
\caption{Non-bulge sample data} \label{tab:nobul_datos} \\
\hline
\hline
No. & Source ID  & $z$ &  Qz &  R (arcsec) & $\epsilon$ & Sample & $I_{814}$  & $M_B$ & $M_R$\\
(1) & (2) & (3) & (4) & (5) & (6) & (7) & (8) & (9) & (10)\\
\hline
\endfirsthead
\caption{Continued.} \\
\hline
\hline
No. & Source ID  & $z$ &  Qz &  R (arcsec) & $\epsilon$ & Sample & $I_{814}$  & $M_B$ & $M_R$\\
(1) & (2) & (3) & (4) & (5) & (6) & (7) & (8) & (9) & (10)\\
\hline
\endhead
\hline
\hline
\endfoot
   86  & 141609.29+521058.8	 & 	0.035	 &   2   &    1.73    &      0.46 &  hi  &     $21.12 \pm 0.02$       &      -13.57  &     -14.37     \\
 1098  & 141633.59+521553.5	 & 	 0.04	 &   2   &    2.02    &      0.16 &  li  &     $18.20 \pm 0.01$       &      -16.72  &     -17.57     \\
  133  & 141733.11+522825.1	 &     0.0667	 &   1   &    2.17    &      0.49 &  hi  &     $20.78 \pm 0.02$       &      -15.28  &     -16.20     \\
  164  & 141725.18+522716.7	 &     0.0723	 &   1   &    3.30    &      0.47 &  hi  &     $20.62 \pm 0.02$       &      -15.82  &     -16.53     \\
  165  & 141525.60+520334.0	 &     0.0728	 &   1   &    3.38    &      0.59 &  hi  &     $19.46 \pm 0.01$       &      -16.71  &     -17.66     \\
  138  & 141740.55+522619.8	 & 	 0.08	 &   1   &    2.24    &      0.58 &  hi  &     $21.07 \pm 0.02$       &      -15.81  &     -16.36     \\
  131  & 141538.06+520529.0	 &     0.0866	 &   1   &    2.17    &      0.48 &  hi  &     $19.84 \pm 0.01$       &      -16.91  &     -17.72     \\
  148  & 141652.56+522049.9	 & 	0.105	 &   2   &    2.52    &      0.50 &  hi  &     $21.03 \pm 0.02$       &      -16.33  &     -16.99     \\
\end{longtable}}

\longtab{3}{
\begin{longtable}[!hbp]{r c c c r r r c c c}
\caption{Bulge sample colours} \label{tab:bul_colours}  \\
\hline
\hline
\multicolumn{1}{c}{No.} & \multicolumn{1}{c}{$(V_{606}-I_{814})_{\rm b}$}  & \multicolumn{1}{c}{$(V_{606}-I_{814})_{\rm d}$}  & \multicolumn{1}{c}{$(V_{606}-I_{814})_{\rm g}$} & \multicolumn{1}{c}{$(U-B)_{\rm b}$}  & \multicolumn{1}{c}{$(U-B)_{\rm d}$}  & \multicolumn{1}{c}{$(U-B)_{\rm g}$} & \multicolumn{1}{c}{$(B-R)_{\rm b}$}  & \multicolumn{1}{c}{$(B-R)_{\rm d}$}  & \multicolumn{1}{c}{$(B-R)_{\rm g}$}  \\
 \multicolumn{1}{c}{(1)} & \multicolumn{1}{c}{(2)} & \multicolumn{1}{c}{(3)} & \multicolumn{1}{c}{(4)} & \multicolumn{1}{c}{(5)} & \multicolumn{1}{c}{(6)} & \multicolumn{1}{c}{(7)} & \multicolumn{1}{c}{(8)} & \multicolumn{1}{c}{(9)} & \multicolumn{1}{c}{(10)} \\
\hline
\endfirsthead
\caption{Continued.} \\
\hline
\hline
\multicolumn{1}{c}{No.} & \multicolumn{1}{c}{$(V_{606}-I_{814})_{\rm b}$}  & \multicolumn{1}{c}{$(V_{606}-I_{814})_{\rm d}$}  & \multicolumn{1}{c}{$(V_{606}-I_{814})_{\rm g}$} & \multicolumn{1}{c}{$(U-B)_{\rm b}$}  & \multicolumn{1}{c}{$(U-B)_{\rm d}$}  & \multicolumn{1}{c}{$(U-B)_{\rm g}$} & \multicolumn{1}{c}{$(B-R)_{\rm b}$}  & \multicolumn{1}{c}{$(B-R)_{\rm d}$}  & \multicolumn{1}{c}{$(B-R)_{\rm g}$}  \\
 \multicolumn{1}{c}{(1)} & \multicolumn{1}{c}{(2)} & \multicolumn{1}{c}{(3)} & \multicolumn{1}{c}{(4)} & \multicolumn{1}{c}{(5)} & \multicolumn{1}{c}{(6)} & \multicolumn{1}{c}{(7)} & \multicolumn{1}{c}{(8)} & \multicolumn{1}{c}{(9)} & \multicolumn{1}{c}{(10)} \\
\hline
\endhead
\hline
\hline
\endfoot
 1131  &   $0.67 \pm 0.33$    &   $0.75 \pm 0.03$   &	$0.64 \pm 0.00$    &	  -0.13  &  -0.06   &	-0.17   &   0.79  &  0.87     &    0.76   \\
 1139  &   $0.92 \pm 0.20$    &   $0.95 \pm 0.05$   &	$0.93 \pm 0.00$    &	   0.42  &   0.45   &	 0.43   &   1.37  &  1.40     &    1.38   \\
 1078  &   $0.76 \pm 0.12$    &   $0.86 \pm 0.04$   &	$0.81 \pm 0.01$    &	   0.08  &   0.18   &	 0.14   &   1.03  &  1.13     &    1.08   \\
 1081  &   $1.08 \pm 0.26$    &   $1.03 \pm 0.18$   &	$1.06 \pm 0.00$    &	   0.54  &   0.49   &	 0.52   &   1.49  &  1.45     &    1.47   \\
 1125  &   $1.00 \pm 0.45$    &   $1.00 \pm 0.11$   &	$1.08 \pm 0.01$    &	   0.27  &   0.27   &	 0.35   &   1.34  &  1.34     &    1.41   \\
  139  &   $0.82 \pm 0.26$    &   $0.74 \pm 1.27$   &	$0.91 \pm 0.02$    &	   0.05  &  -0.04   &	 0.14   &   1.01  &  0.92     &    1.10   \\
  170  &   $0.95 \pm 0.13$    &   $0.66 \pm 0.15$   &	$0.79 \pm 0.01$    &	   0.12  &  -0.17   &	-0.04   &   1.05  &  0.77     &    0.90   \\
  159  &   $1.16 \pm 0.06$    &   $1.09 \pm 0.11$   &	$1.16 \pm 0.00$    &	   0.41  &   0.33   &	 0.41   &   1.41  &  1.34     &    1.41   \\
\end{longtable}}

\longtab{4}{
\begin{longtable}[!hbp]{r c c c r r r c c c}
\caption{Non-bulge sample colours}\label{tab:nobul_colours}  \\
\hline
\hline
\multicolumn{1}{c}{No.} & \multicolumn{1}{c}{$(V_{606}-I_{814})_{\rm b}$}  & \multicolumn{1}{c}{$(V_{606}-I_{814})_{\rm d}$}  & \multicolumn{1}{c}{$(V_{606}-I_{814})_{\rm g}$} & \multicolumn{1}{c}{$(U-B)_{\rm b}$}  & \multicolumn{1}{c}{$(U-B)_{\rm d}$}  & \multicolumn{1}{c}{$(U-B)_{\rm g}$} & \multicolumn{1}{c}{$(B-R)_{\rm b}$}  & \multicolumn{1}{c}{$(B-R)_{\rm d}$}  & \multicolumn{1}{c}{$(B-R)_{\rm g}$}  \\
 \multicolumn{1}{c}{(1)} & \multicolumn{1}{c}{(2)} & \multicolumn{1}{c}{(3)} & \multicolumn{1}{c}{(4)} & \multicolumn{1}{c}{(5)} & \multicolumn{1}{c}{(6)} & \multicolumn{1}{c}{(7)} & \multicolumn{1}{c}{(8)} & \multicolumn{1}{c}{(9)} & \multicolumn{1}{c}{(10)} \\
\hline
\endfirsthead
\caption{Continued.} \\
\hline
\hline
\multicolumn{1}{c}{No.} & \multicolumn{1}{c}{$(V_{606}-I_{814})_{\rm b}$}  & \multicolumn{1}{c}{$(V_{606}-I_{814})_{\rm d}$}  & \multicolumn{1}{c}{$(V_{606}-I_{814})_{\rm g}$} & \multicolumn{1}{c}{$(U-B)_{\rm b}$}  & \multicolumn{1}{c}{$(U-B)_{\rm d}$}  & \multicolumn{1}{c}{$(U-B)_{\rm g}$} & \multicolumn{1}{c}{$(B-R)_{\rm b}$}  & \multicolumn{1}{c}{$(B-R)_{\rm d}$}  & \multicolumn{1}{c}{$(B-R)_{\rm g}$}  \\
 \multicolumn{1}{c}{(1)} & \multicolumn{1}{c}{(2)} & \multicolumn{1}{c}{(3)} & \multicolumn{1}{c}{(4)} & \multicolumn{1}{c}{(5)} & \multicolumn{1}{c}{(6)} & \multicolumn{1}{c}{(7)} & \multicolumn{1}{c}{(8)} & \multicolumn{1}{c}{(9)} & \multicolumn{1}{c}{(10)} \\
\hline
\endhead
\hline
\hline
\endfoot
   86	 &     $0.75 \pm 3.76$   &   $0.78 \pm 0.10$   &     $0.67 \pm 0.02$  	&     -0.05   &  -0.01     &	   -0.12  &	0.90   &   0.93    &	   0.82   \\
 1098	 &     $0.83 \pm 0.18$   &   $0.77 \pm 0.03$   &     $0.70 \pm 0.00$  	&      0.00   &  -0.05     &	   -0.13  &	0.99   &   0.93    &	   0.86   \\
  133	 &     $0.50 \pm 0.53$   &   $0.75 \pm 0.06$   &     $0.74 \pm 0.02$  	&      0.13   &   0.38     &	    0.37  &	0.81   &   1.06    &	   1.05   \\
  164	 &     $0.76 \pm 0.33$   &   $0.79 \pm 0.24$   &     $0.63 \pm 0.02$  	&      0.07   &   0.10     &	   -0.06  &	0.86   &   0.89    &	   0.74   \\
  165	 &     $0.72 \pm 0.23$   &   $0.75 \pm 0.06$   &     $0.74 \pm 0.01$  	&      0.08   &   0.11     &	    0.10  &	0.98   &   1.01    &	   1.00   \\
  138	 &     $0.71 \pm 0.43$   &   $0.69 \pm 0.19$   &     $0.51 \pm 0.02$  	&      0.06   &   0.05     &	   -0.13  &	0.77   &   0.75    &	   0.57   \\
  131	 &     $0.70 \pm 0.11$   &   $0.73 \pm 0.05$   &     $0.71 \pm 0.01$  	&      0.08   &   0.10     &	    0.08  &	0.91   &   0.93    &	   0.92   \\
  148	 &     $0.97 \pm 0.33$   &   $0.71 \pm 0.25$   &     $0.63 \pm 0.02$  	&      0.36   &   0.10     &	    0.01  &	1.08   &   0.83    &	   0.74   \\
\end{longtable}}

\longtab{5}{
\begin{longtable}[!hbp]{r r r r}
\caption{Bulge sample colour gradients} \label{tab:bul_gradient} \\
\hline
\hline
\multicolumn{1}{c}{No.} &  \multicolumn{1}{c}{$\bigtriangledown(V_{606}-I_{814})_{\rm blue}$} & \multicolumn{1}{c}{$\bigtriangledown(V_{606}-I_{814})_{\rm red}$} & \multicolumn{1}{c}{$\bigtriangledown(V_{606}-I_{814})_{\rm sma}$}     \\
\multicolumn{1}{c}{(1)} & \multicolumn{1}{c}{(2)} & \multicolumn{1}{c}{(3)} & \multicolumn{1}{c}{(4)} \\
\hline
\endfirsthead
\caption{Continued.} \\
\hline
\hline
\multicolumn{1}{c}{No.} &  \multicolumn{1}{c}{$\bigtriangledown(V_{606}-I_{814})_{\rm blue}$} & \multicolumn{1}{c}{$\bigtriangledown(V_{606}-I_{814})_{\rm red}$} & \multicolumn{1}{c}{$\bigtriangledown(V_{606}-I_{814})_{\rm sma}$}     \\
\multicolumn{1}{c}{(1)} & \multicolumn{1}{c}{(2)} & \multicolumn{1}{c}{(3)} & \multicolumn{1}{c}{(4)} \\
\hline
\endhead
\hline
\hline
\endfoot
 1131 &  $0.102 \pm 0.020$ & -$0.014 \pm 0.051$ &  $0.066 \pm 0.010$   \\
 1139 & -$0.005 \pm 0.003$ & -$0.011 \pm 0.004$ &  $0.022 \pm 0.005$	\\
 1078 & -$0.017 \pm 0.021$ & -$0.028 \pm 0.023$ & -$0.014 \pm 0.018$   \\
 1081 & -$0.030 \pm 0.015$ & -$0.049 \pm 0.012$ & -$0.012 \pm 0.013$   \\
 1125 & -$0.024 \pm 0.010$ & -$0.034 \pm 0.008$ & -$0.038 \pm 0.008$	\\
  139 & -$0.045 \pm 0.010$ & -$0.038 \pm 0.018$ & -$0.045 \pm 0.018$   \\
  170 & -$0.035 \pm 0.004$ & -$0.021 \pm 0.004$ & -$0.015 \pm 0.005$	\\
  159 & -$0.013 \pm 0.005$ & -$0.020 \pm 0.005$ & -$0.017 \pm 0.005$	\\
\end{longtable}}

\longtab{6}{
\begin{longtable}[!hbp]{r r r r}
\caption{Non-bulge sample colour gradients} \label{tab:nobul_gradient} \\
\hline
\hline
\multicolumn{1}{c}{No.} &  \multicolumn{1}{c}{$\bigtriangledown(V_{606}-I_{814})_{\rm blue}$} & \multicolumn{1}{c}{$\bigtriangledown(V_{606}-I_{814})_{\rm red}$} & \multicolumn{1}{c}{$\bigtriangledown(V_{606}-I_{814})_{\rm sma}$}     \\
\multicolumn{1}{c}{(1)} & \multicolumn{1}{c}{(2)} & \multicolumn{1}{c}{(3)} & \multicolumn{1}{c}{(4)} \\
\hline
\endfirsthead
\caption{Continued.} \\
\hline
\hline
\multicolumn{1}{c}{No.} &  \multicolumn{1}{c}{$\bigtriangledown(V_{606}-I_{814})_{\rm blue}$} & \multicolumn{1}{c}{$\bigtriangledown(V_{606}-I_{814})_{\rm red}$} & \multicolumn{1}{c}{$\bigtriangledown(V_{606}-I_{814})_{\rm sma}$}     \\
\multicolumn{1}{c}{(1)} & \multicolumn{1}{c}{(2)} & \multicolumn{1}{c}{(3)} & \multicolumn{1}{c}{(4)} \\
\hline
\endhead
\hline
\hline
\endfoot
   86	 &      $0.054 \pm 0.142$   &	 -$0.163 \pm 0.180$   &     $0.331 \pm 0.093$	  \\
 1098	 &      $0.039 \pm 0.044$   &	  $0.038 \pm 0.019$   &     $0.069 \pm 0.025$	  \\
  133	 &      $0.053 \pm 0.095$   &	 -$0.026 \pm 0.051$   &    -$0.031 \pm 0.026$	  \\
  164	 &      $0.034 \pm 0.033$   &	 -$0.029 \pm 0.033$   &     $0.020 \pm 0.033$	  \\
  165	 &      $0.037 \pm 0.018$   &	 -$0.011 \pm 0.026$   &    -$0.029 \pm 0.010$	  \\
  138	 &      $0.085 \pm 0.032$   &	  $0.040 \pm 0.040$   &     $0.217 \pm 0.061$	  \\
  131	 &     -$0.000 \pm 0.038$   &	  $0.013 \pm 0.027$   &     $0.060 \pm 0.011$	  \\
  148	 &      $0.124 \pm 0.052$   &	 -$0.084 \pm 0.062$   &    -$0.029 \pm 0.027$	  \\
\end{longtable}}

\end{appendix}

\pagebreak

\onecolumn

\begin{appendix}
\section{Sample of galaxies with bulges}
\label{sec:samplesbulges}

\noindent Notes for figures:

\noindent Column (1): $10 \times 10$ arcsec postage stamp of the source in $F814W$ band  \\
Column (2): $10 \times 10$ arcsec postage stamp of the observed $(F606W - F814W)$ colour map  \\
Column (3): Photometric spectral energy distribution  \\
Column (4): Averaged surface brightness profile along both semi-major axis for $F814W$ and $F606W$ bands. {\it Solid line}: exponential law fitted to the outer region of the galaxy. {\it Vertical dashed lines}: 1 and 2 times the scale length of the exponential fit; this is the range in which disk colours were measured \\
Column (5): Observed colour profiles, $(F814W -F606W)$, over both  semi-minor axes.  \\

\noindent \footnotesize{Note: On top of each row we list: the source  ID; the source sky coordinates, right ascension and declination (RA, DEC, J2000.0); the source number identification; the source redshift  ($z$), either photometric or spectroscopic; the semi-major axis  (sma); the ellipticity (e); and, the position angle of the bluer semi-minor axis in which we measure nuclear colours, counter clockwise  from the horizontal axis (PA\_bluer).}

\newpage

\begin{figure*}
\begin{center}
\includegraphics[angle=0,width=20cm]{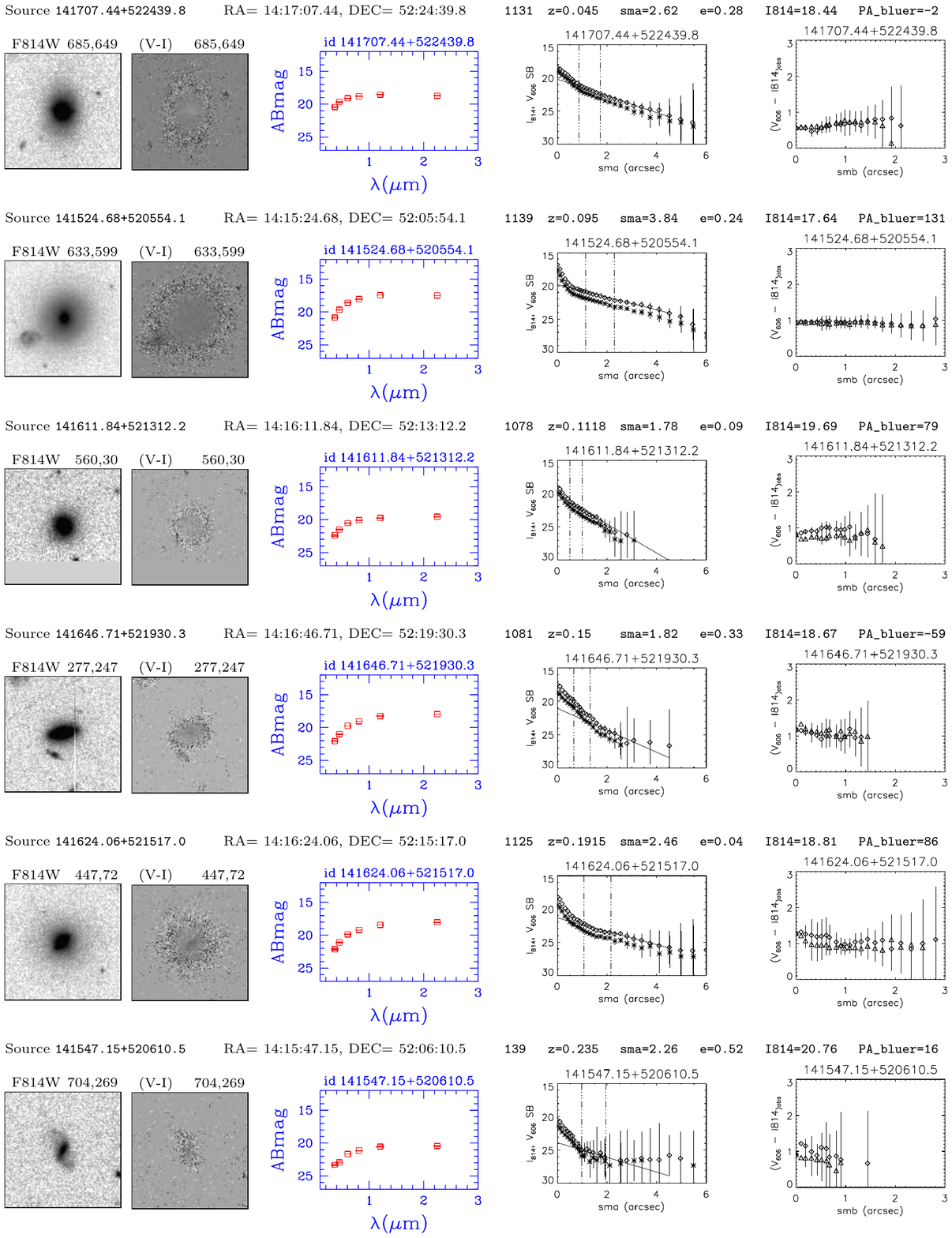}%
\end{center}
\label{fig:pstamps1}
\end{figure*}

\begin{figure*}
\begin{center}
\includegraphics[angle=0,width=20cm]{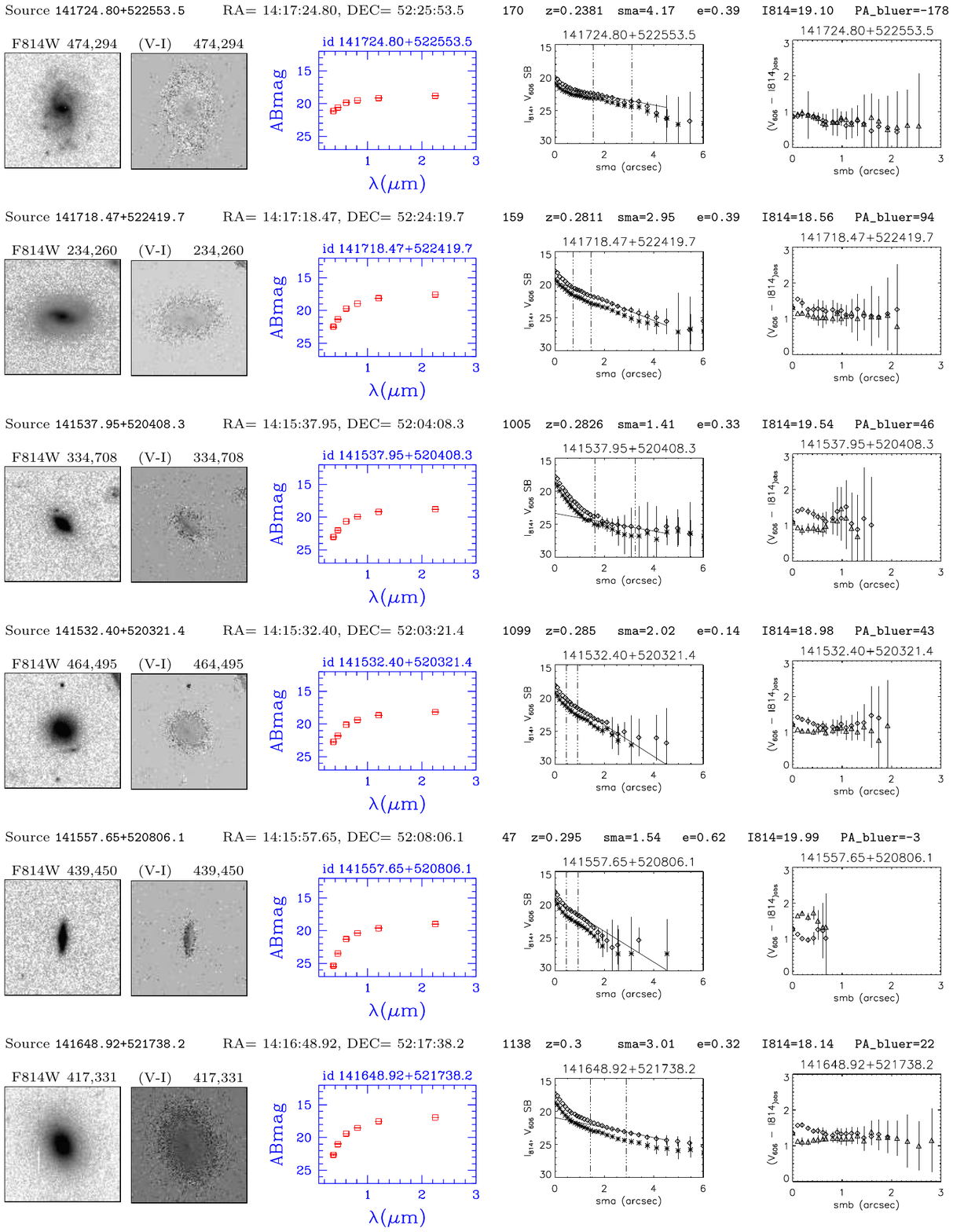}%
\end{center}
\label{fig:pstamps2}
\end{figure*}
\begin{figure*}
\begin{center}
\includegraphics[angle=0,width=20cm]{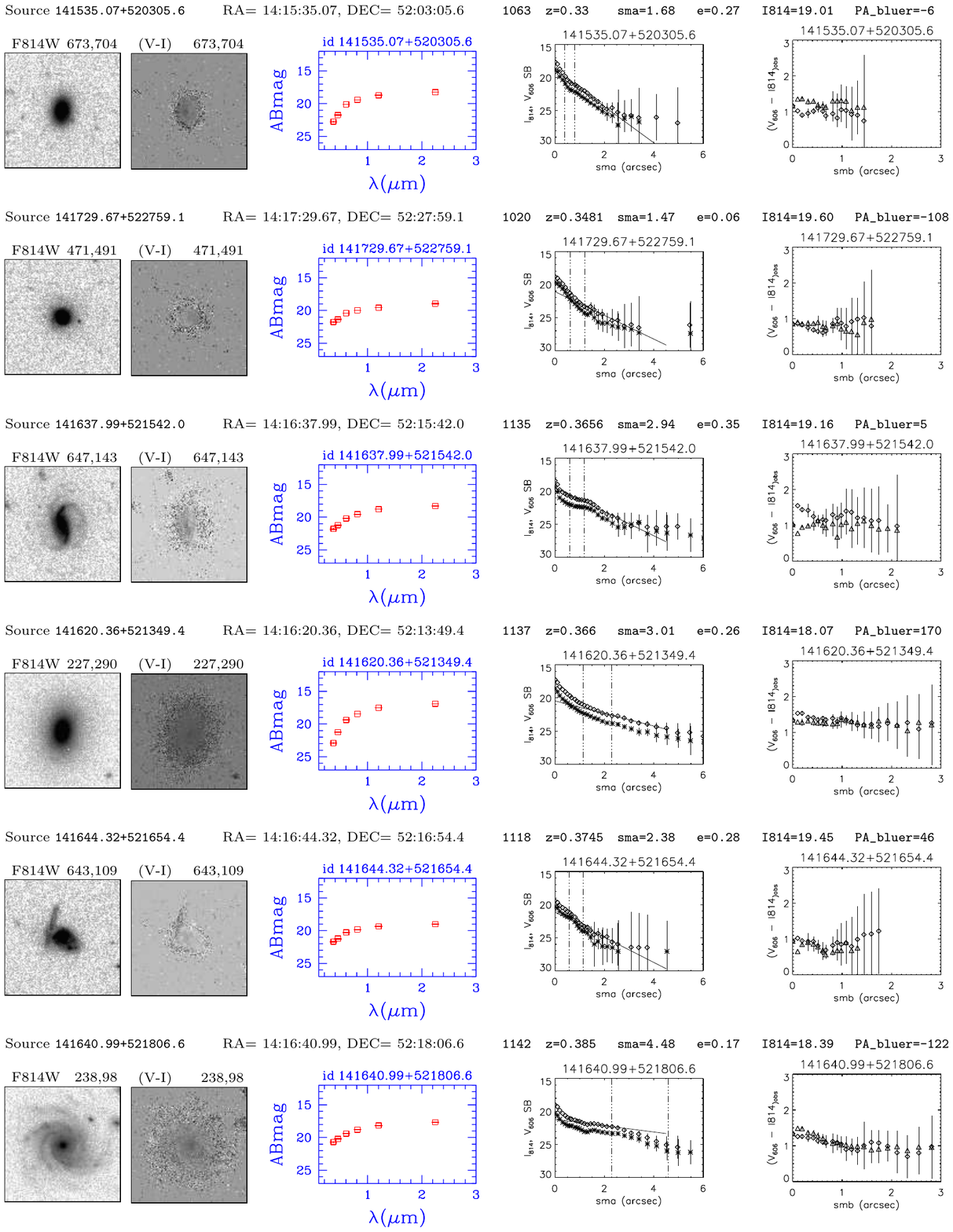}%
\end{center}
\label{fig:pstamps3}
\end{figure*}
\begin{figure*}
\begin{center}
\includegraphics[angle=0,width=20cm]{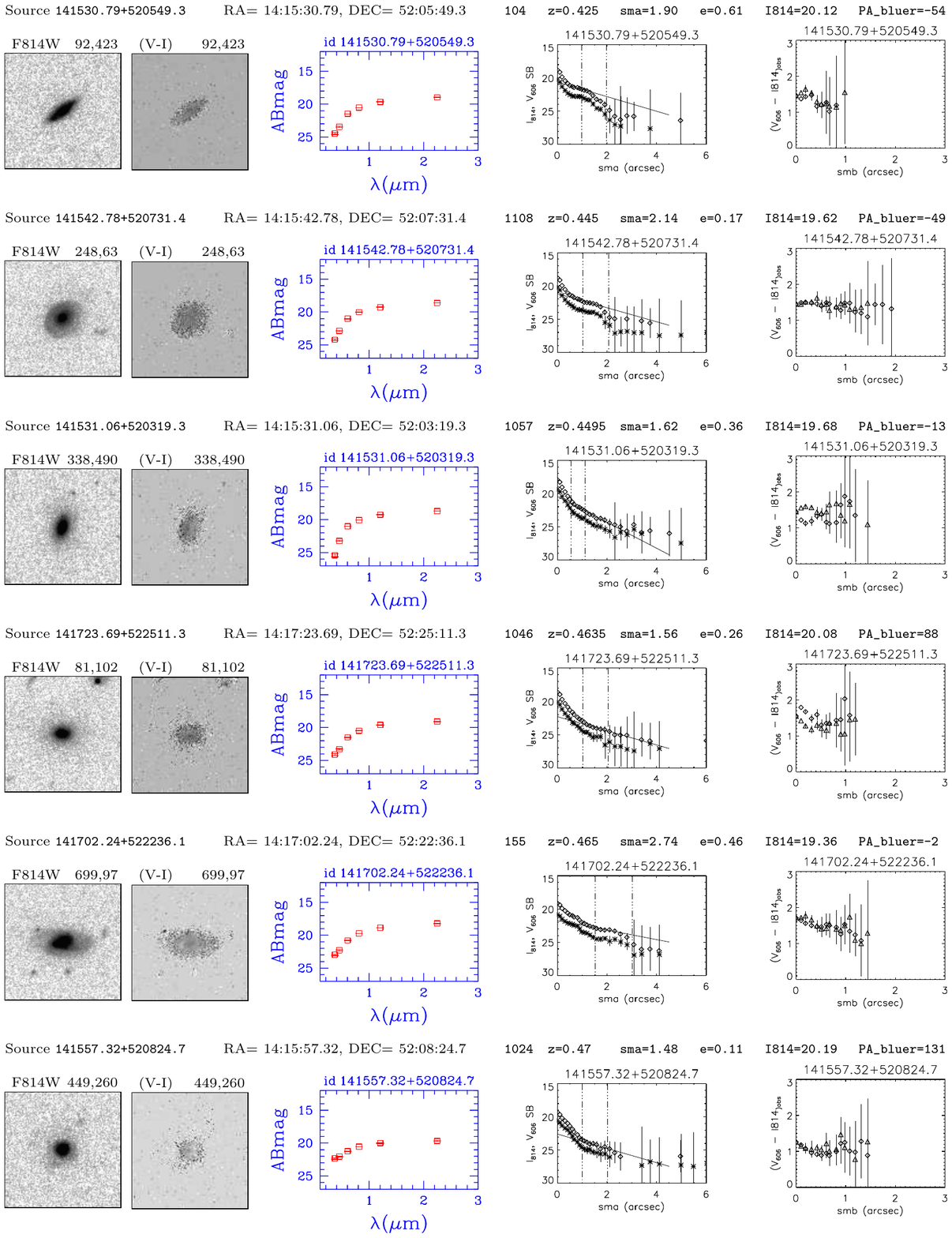}%
\end{center}
\label{fig:pstamps4}
\end{figure*}
\begin{figure*}
\begin{center}
\includegraphics[angle=0,width=20cm]{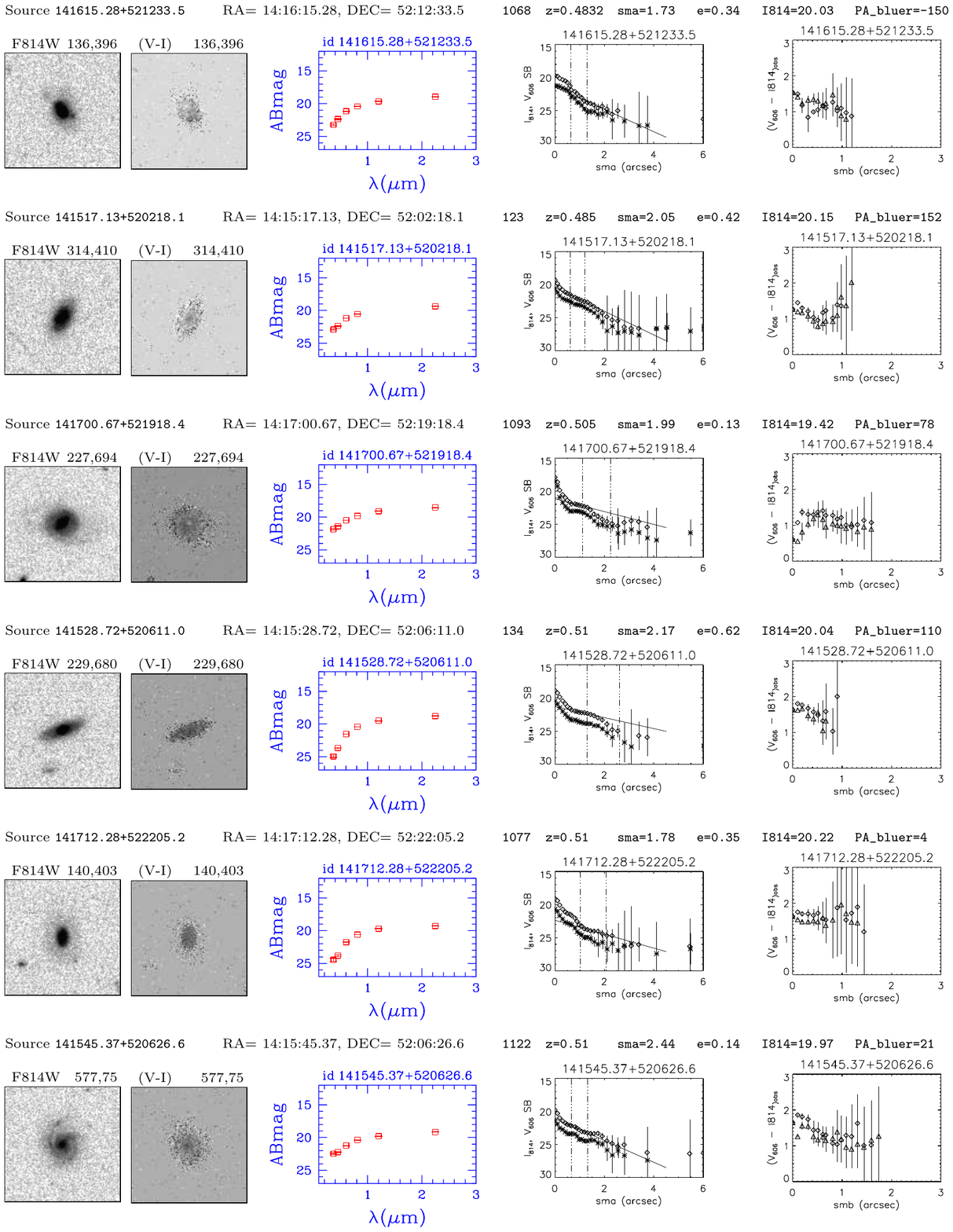}%
\end{center}
\label{fig:pstamps5}
\end{figure*}
\begin{figure*}
\begin{center}
\includegraphics[angle=0,width=20cm]{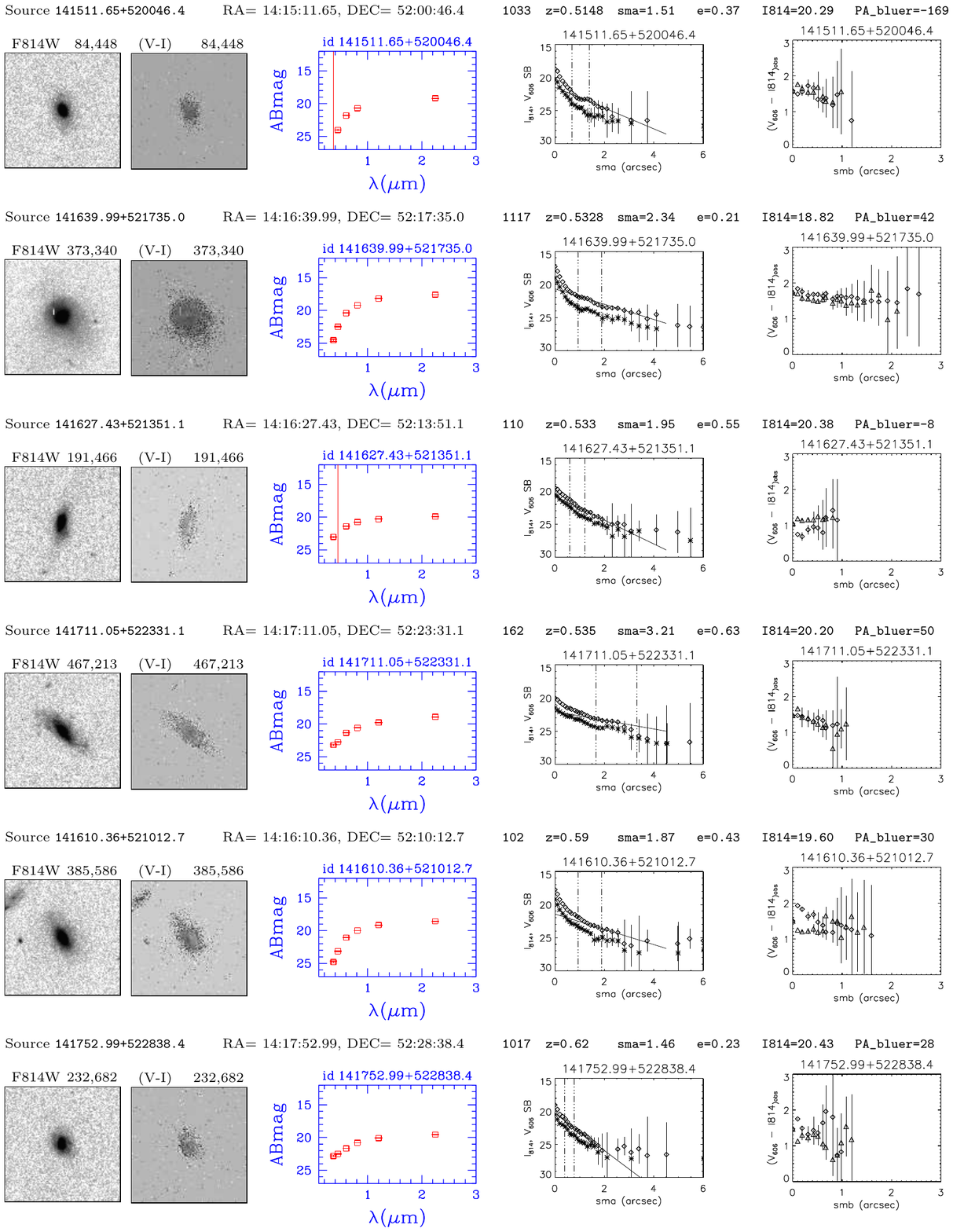}%
\end{center}
\label{fig:pstamps6}
\end{figure*}
\begin{figure*}
\begin{center}
\includegraphics[angle=0,width=20cm]{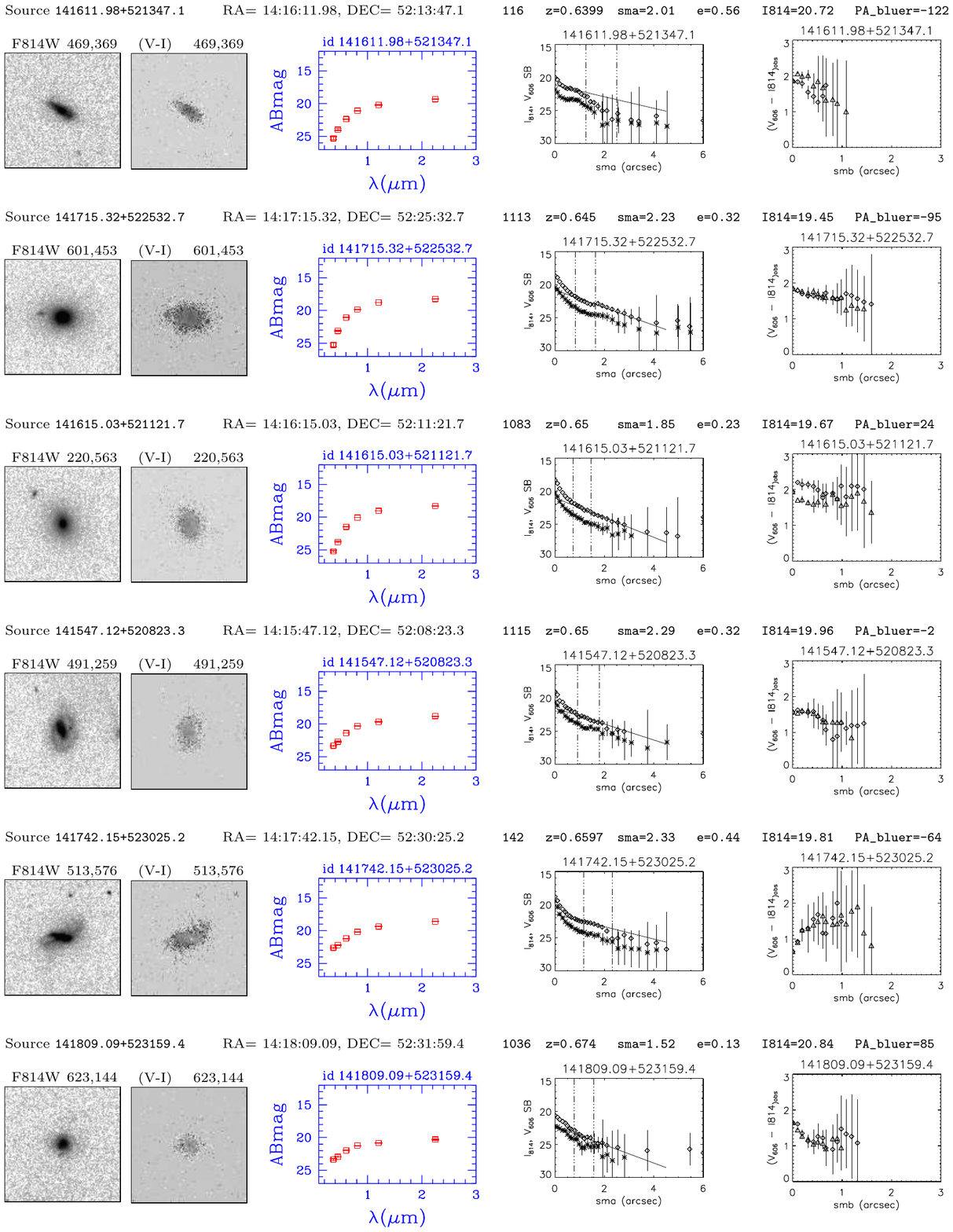}%
\end{center}
\label{fig:pstamps7}
\end{figure*}
\begin{figure*}
\begin{center}
\includegraphics[angle=0,width=20cm]{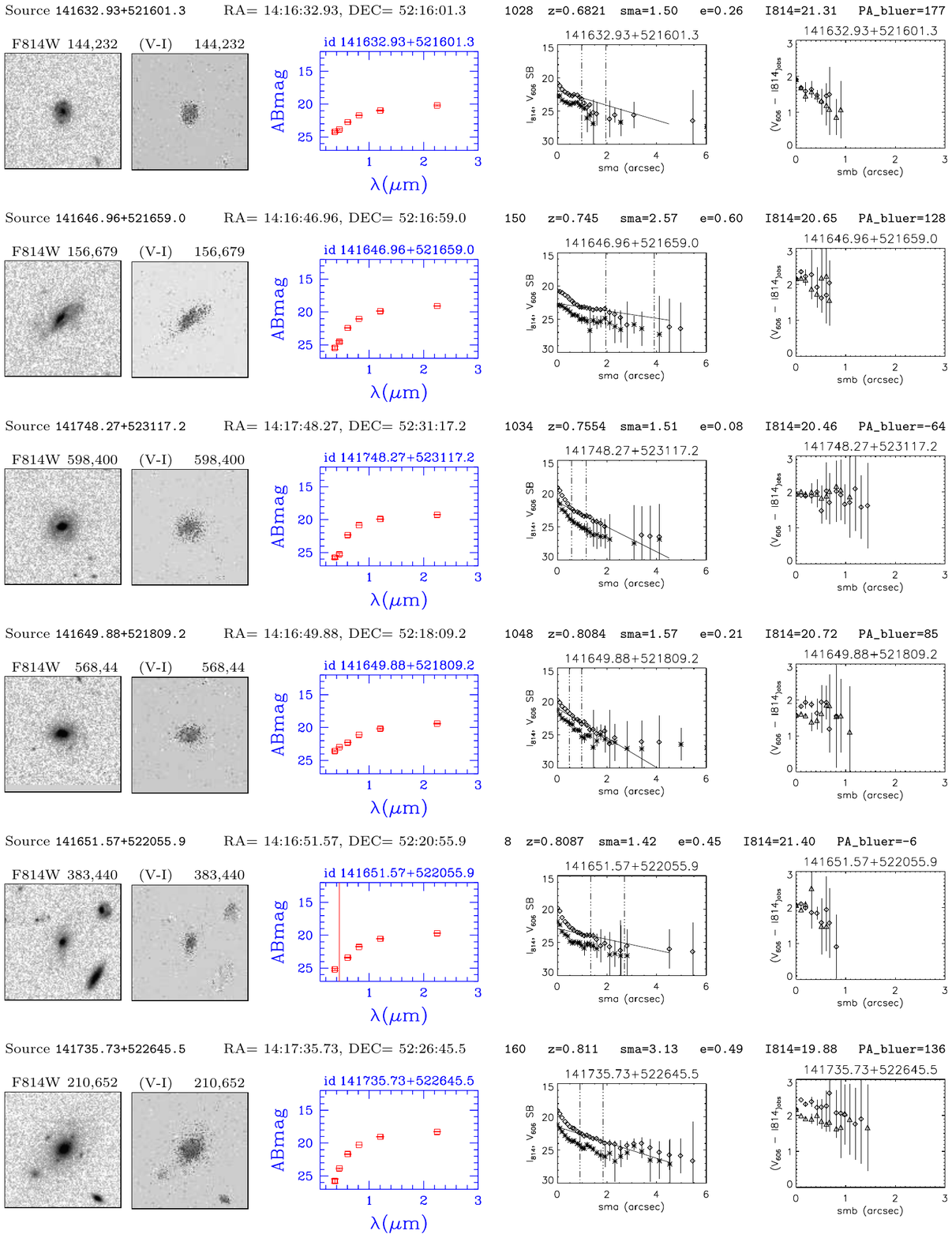}%
\end{center}
\label{fig:pstamps8}
\end{figure*}
\begin{figure*}
\begin{center}
\includegraphics[angle=0,width=20cm]{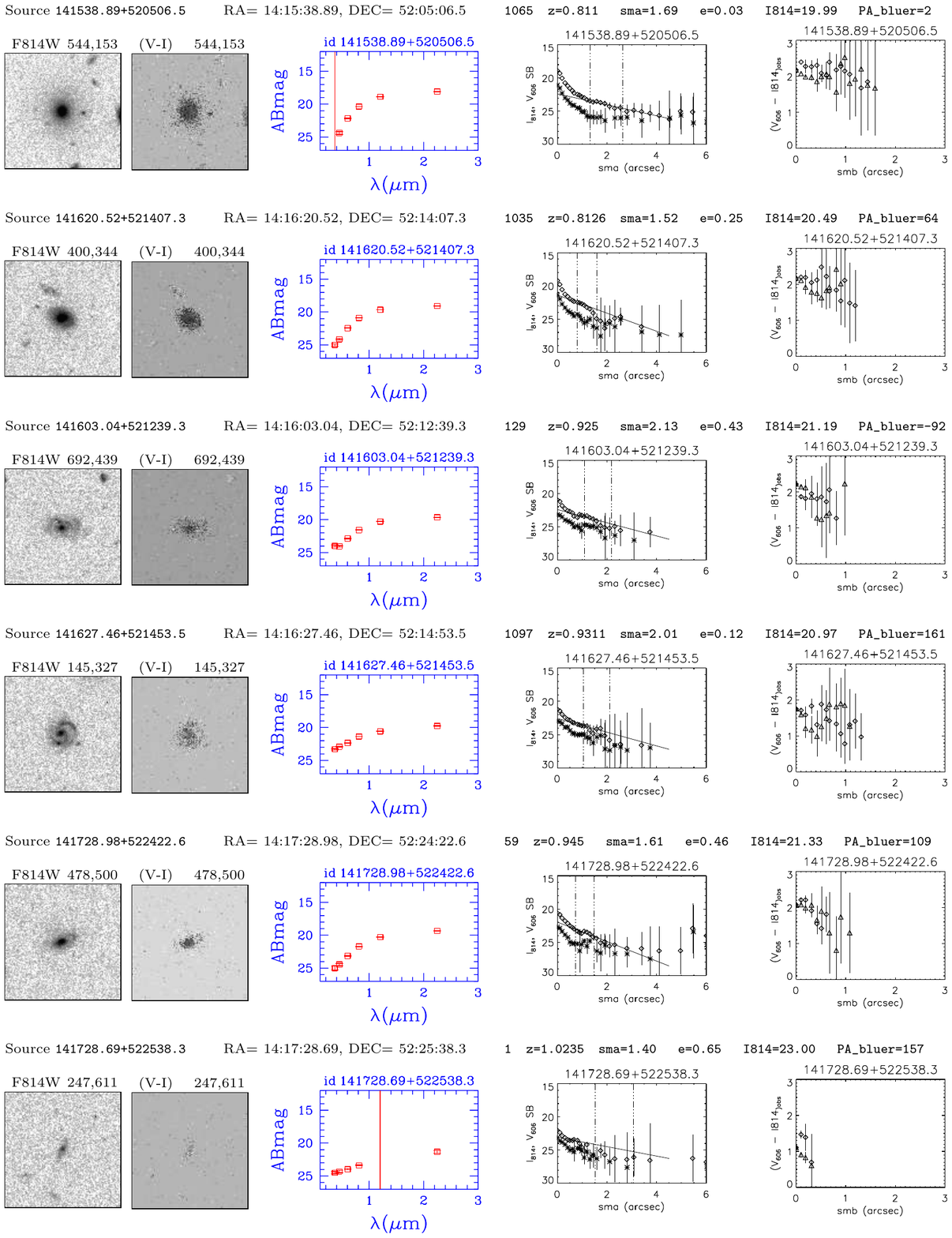}%
\end{center}
\label{fig:pstamps9}
\end{figure*}

\end{appendix}

\onecolumn

\begin{appendix}
\section{Colour profiles in bluer semi-minor axes}
\label{sec:bluercolprof}

\begin{figure*}
\begin{center}
\includegraphics[angle=0,width=7.5cm]{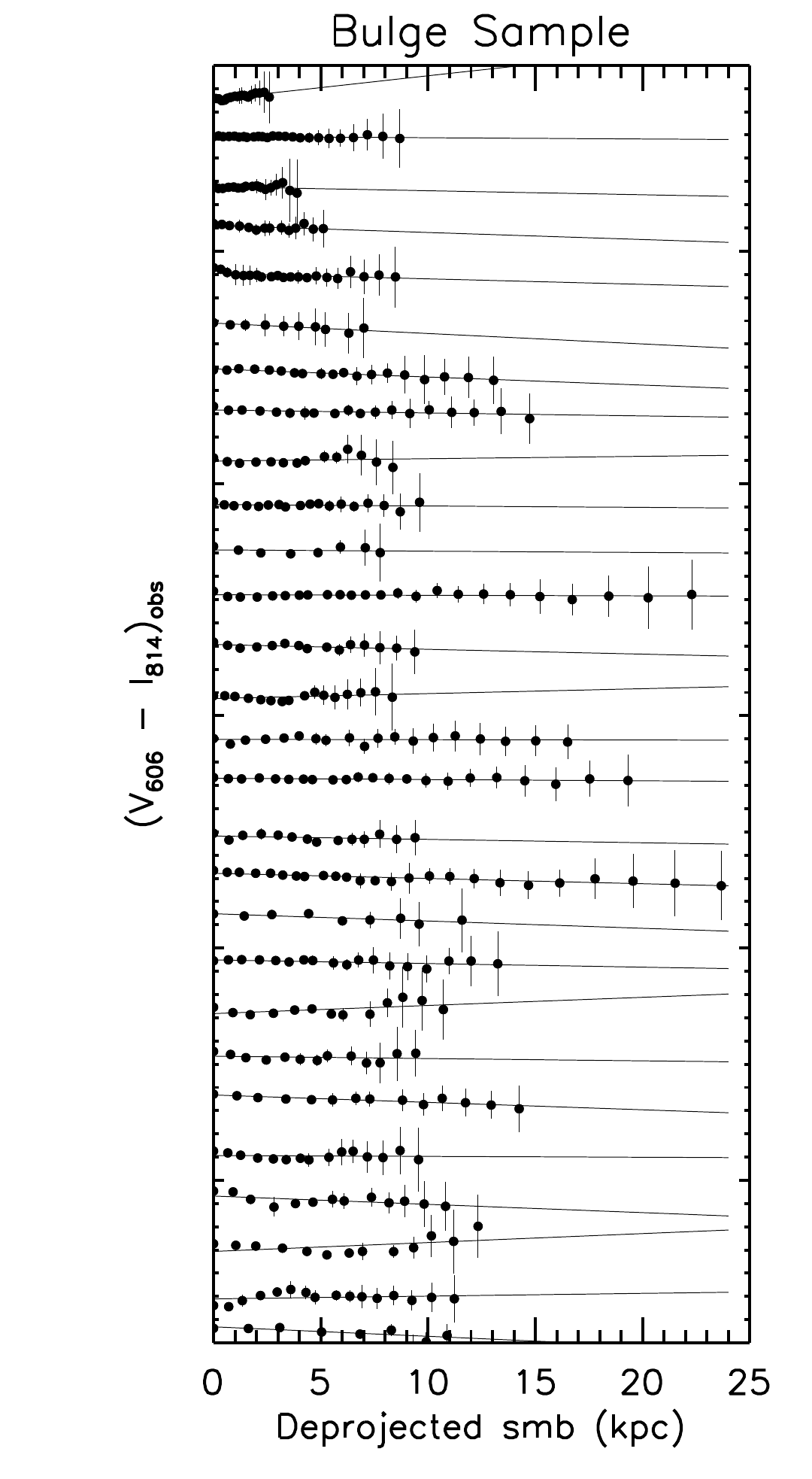}%
\includegraphics[angle=0,width=7.5cm]{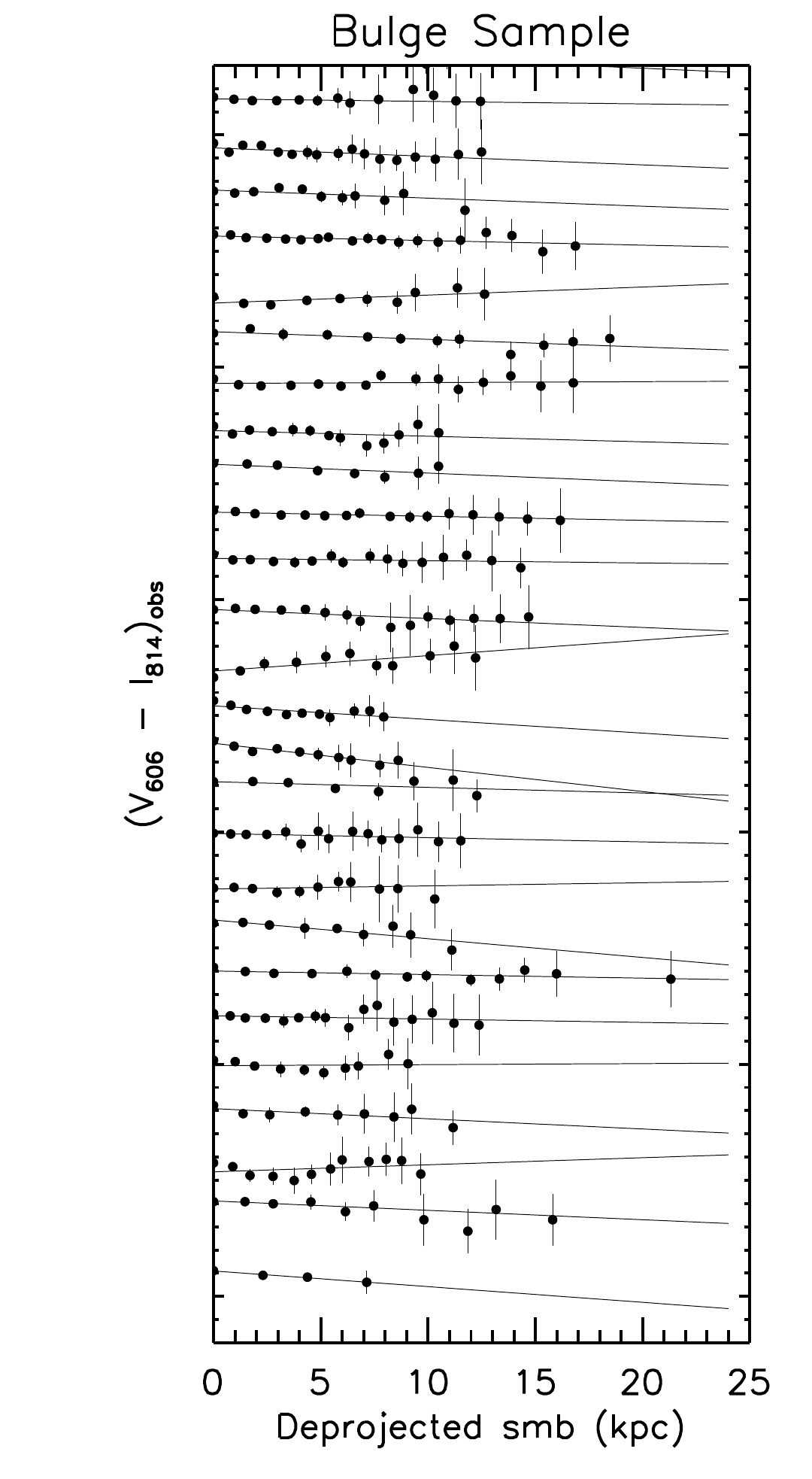}%
\end{center}
\label{fig:colprofbul}
\end{figure*}

\noindent Notes for Figures~\ref{fig:colprofbul}:

\noindent The $(F606W-F814W)$ colour profiles along the bluer deprojected semi-minor axes colour profile, for the bulge and the non-bulge samples.  Overplotted are the linear fits done to calculate the colour  gradients. The profiles are ordered from top to bottom as in the  Appendices~\ref{sec:datatables} and~\ref{sec:samplesbulges}. Minor tick marks on the y-axis correspond to 1 mag in $(F606W-F814W)$. \\

\newpage

\begin{figure*}
\begin{center}
\includegraphics[angle=0,width=7.5cm]{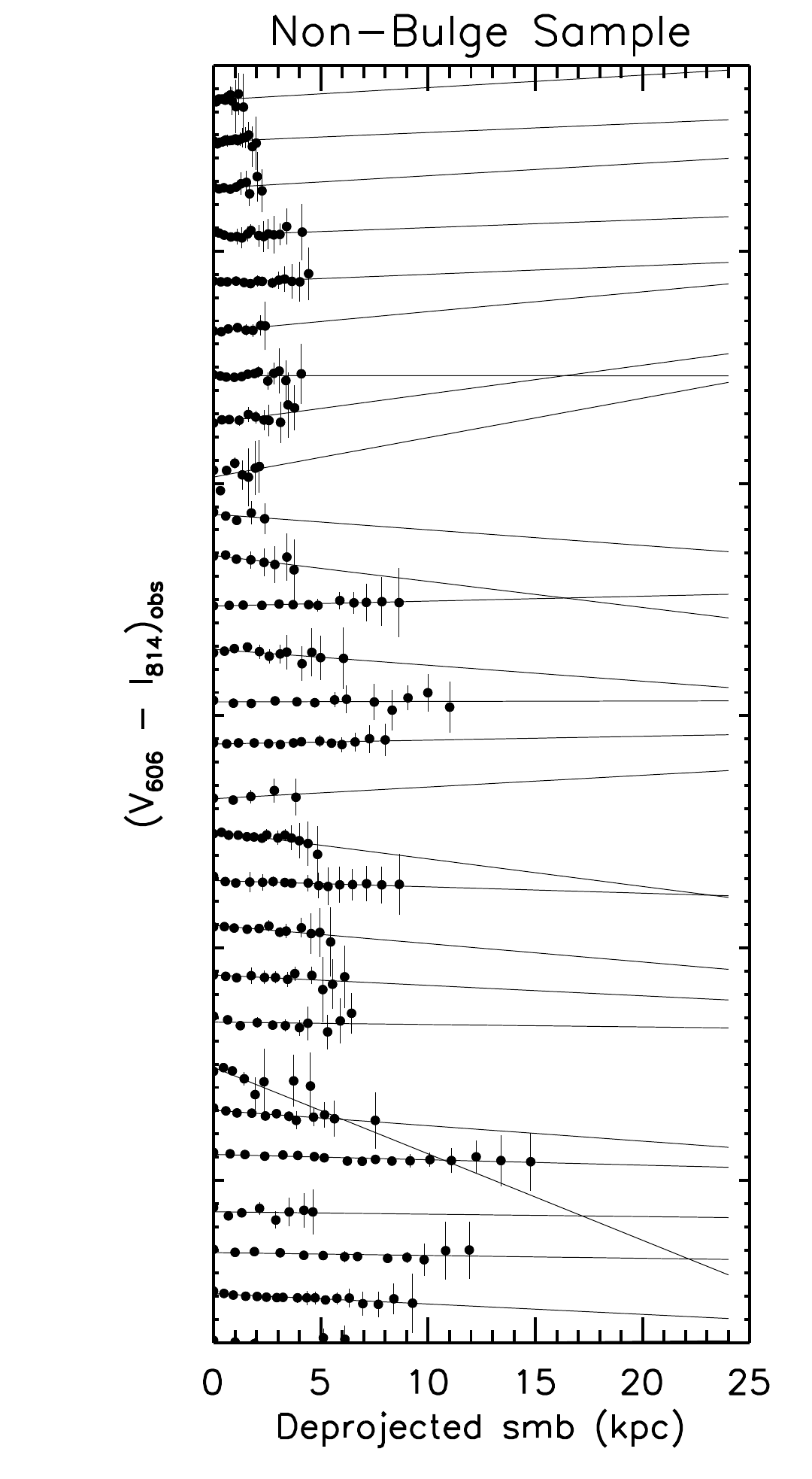}%
\includegraphics[angle=0,width=7.5cm]{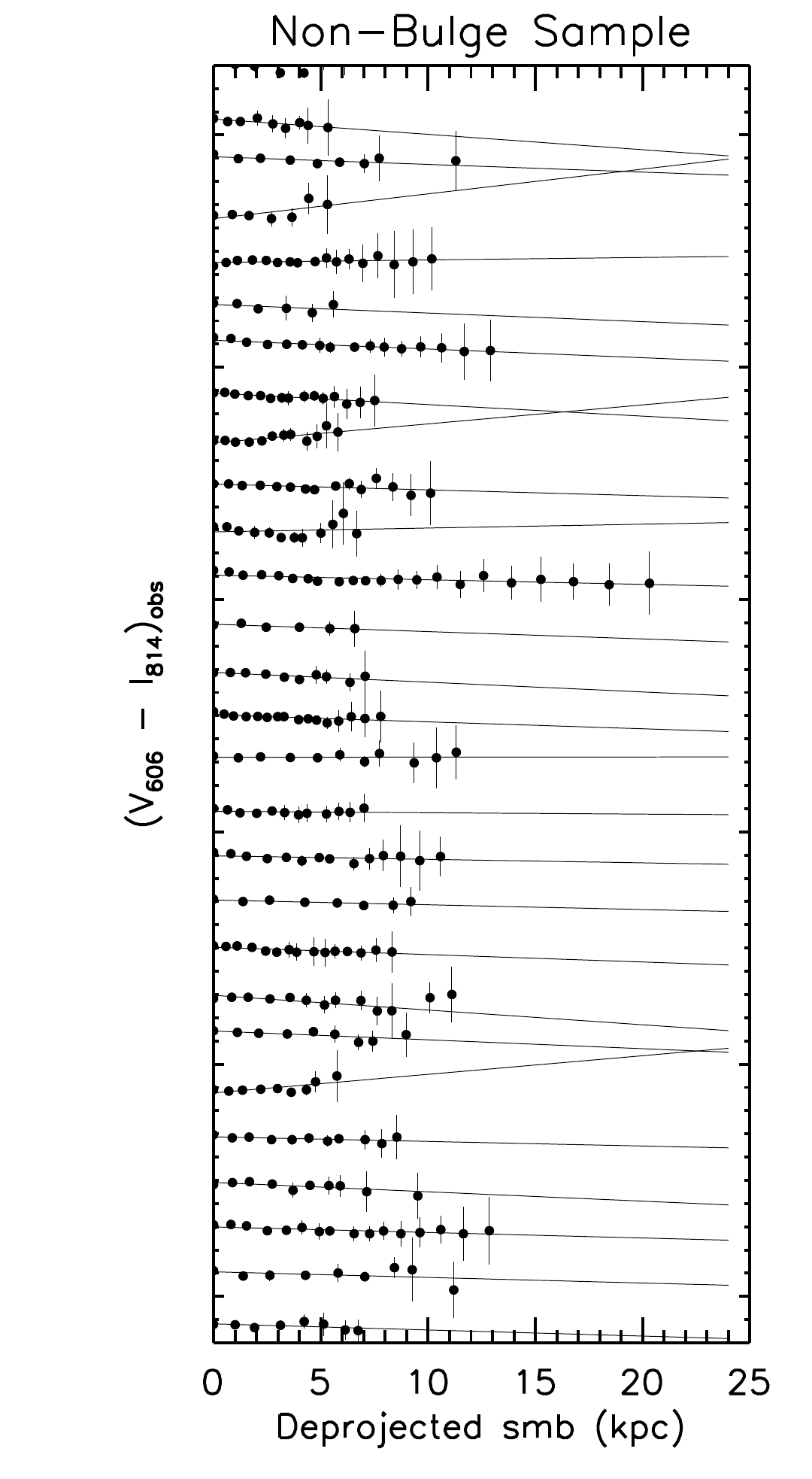}%
\hspace{2in}%
\includegraphics[angle=0,width=7.5cm]{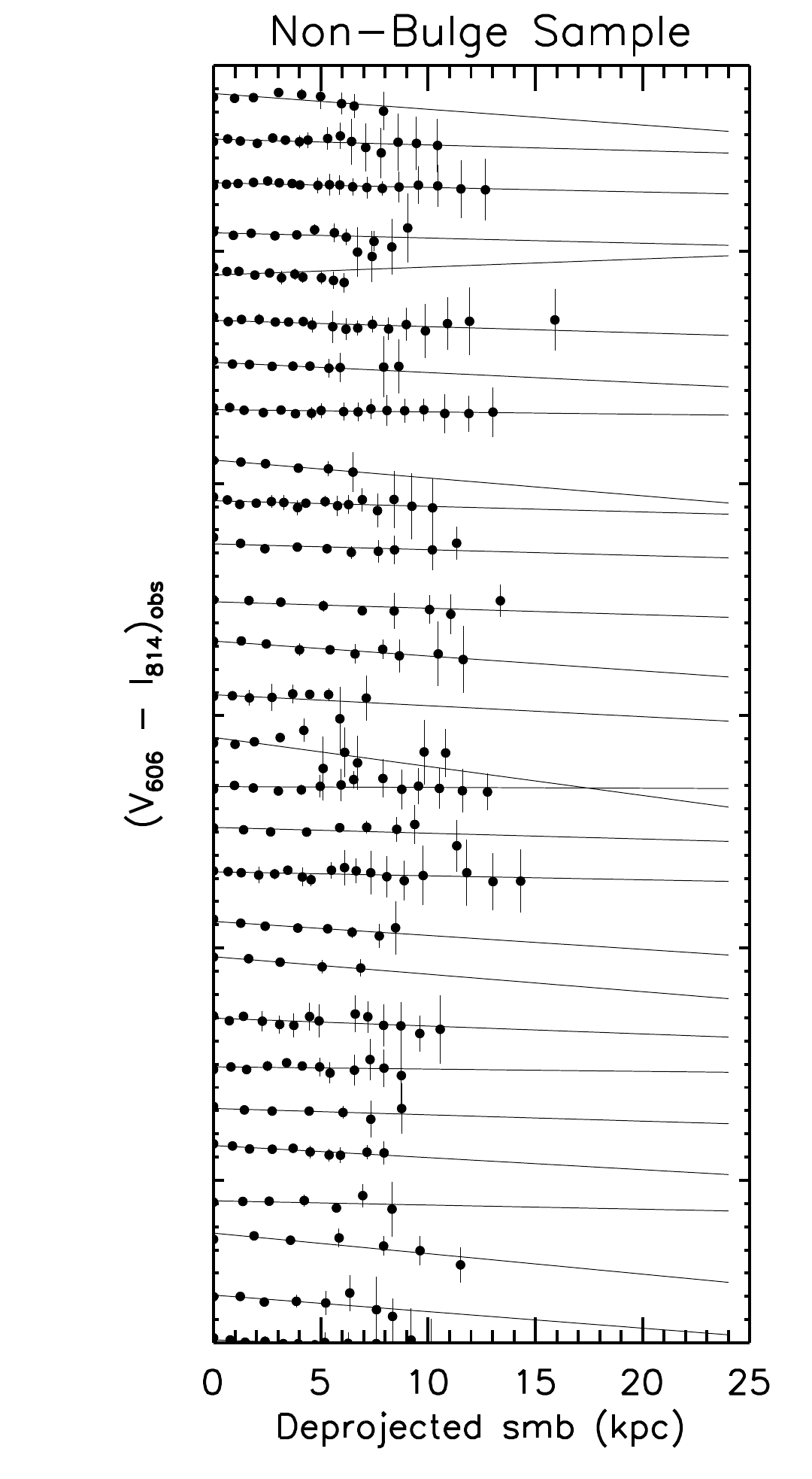}%
\includegraphics[angle=0,width=7.5cm]{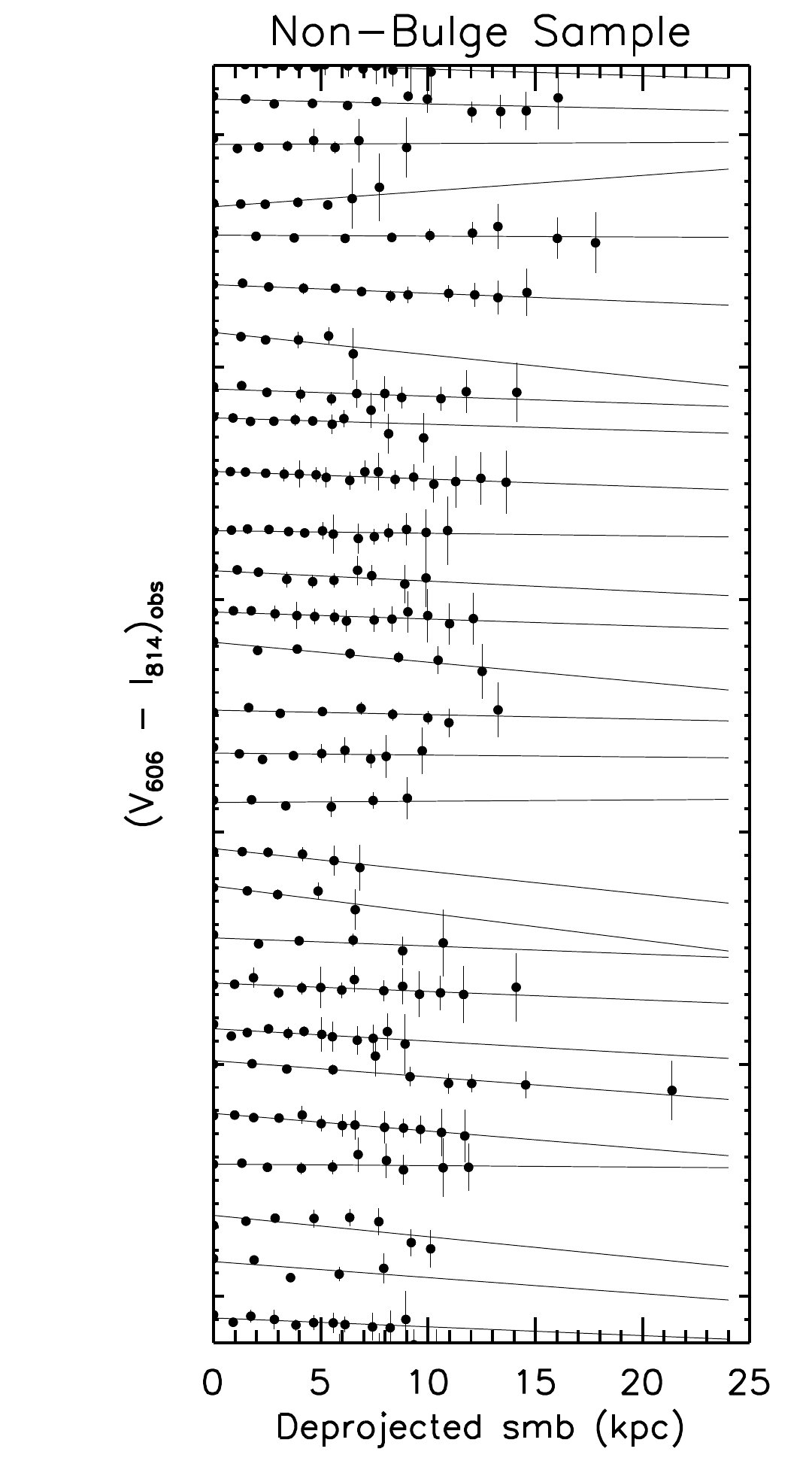}%
\end{center}
\label{fig:colprofnobul}
\end{figure*}

\begin{figure*}
\begin{center}
\includegraphics[angle=0,width=7.5cm]{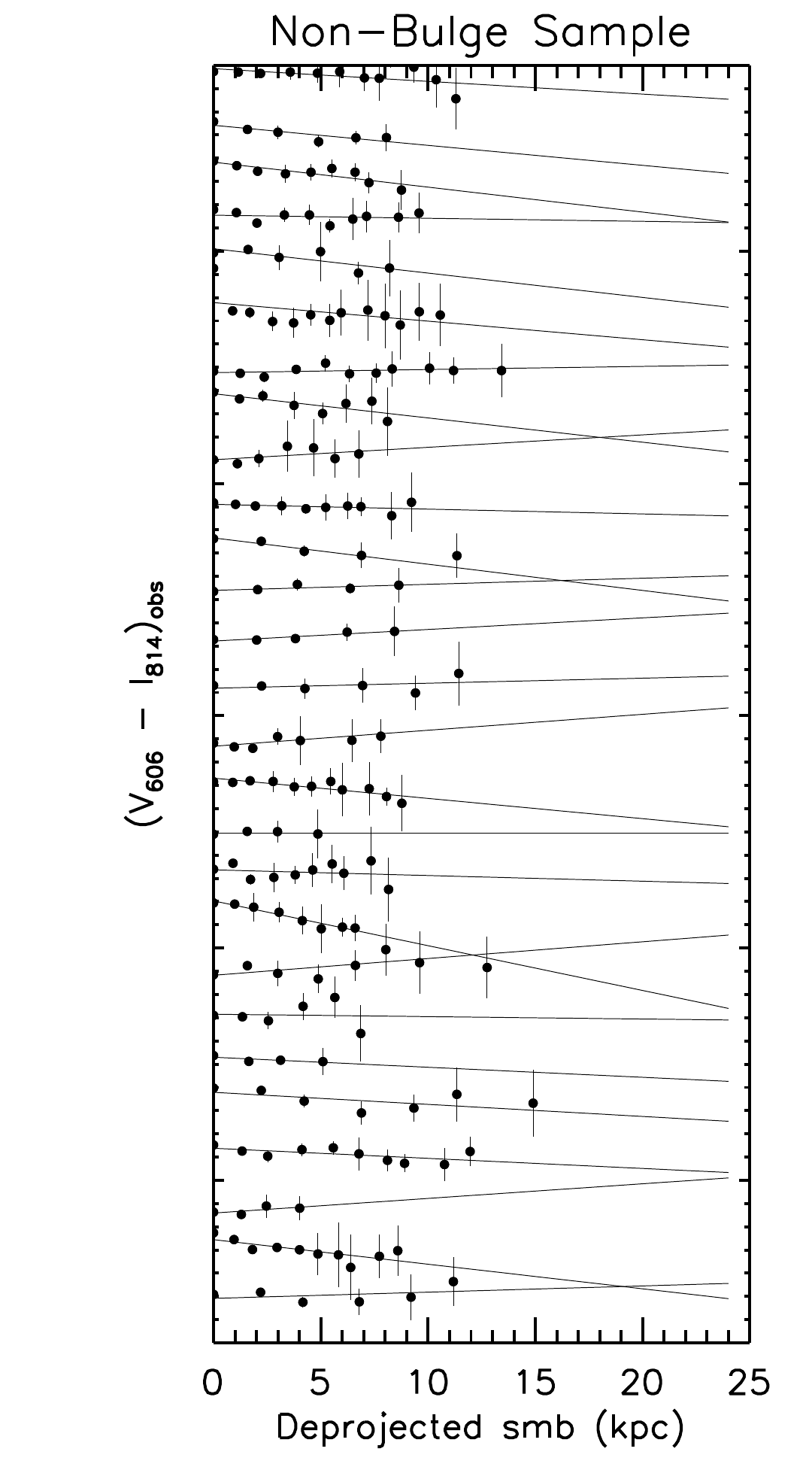}%
\end{center}
\label{fig:colprofnobul}
\end{figure*}

\end{appendix}

\end{document}